\documentclass{aa}
\usepackage{graphicx}
\usepackage{txfonts}
\usepackage{siunitx}
\usepackage{hyperref}
\usepackage{lipsum}
\usepackage{bm}
\usepackage{xcolor}
\usepackage[normalem]{ulem}
\usepackage{subfig}
\usepackage{upgreek}

\newcommand{\rmn}[1]{\mathrm{#1}}
\newcommand{\bnabla}{\bm{\nabla}}
\newcommand{\bcdot}{\bm{\cdot}}
\newcommand{\btimes}{\bm{\times}}

\newcommand{\uproman}[1]{\uppercase\expandafter{\romannumeral#1}}

\usepackage{xpatch}
\usepackage{hyperref}
\hypersetup{
	colorlinks=true,
	breaklinks=true,
	citecolor=blue,
	allcolors=blue,
	frenchlinks=true
}

\makeatletter
\xpatchcmd\NAT@citex
{%
	\@citea\NAT@hyper@{%
		\NAT@nmfmt{\NAT@nm}%
		\hyper@natlinkbreak{\NAT@aysep\NAT@spacechar}{\@citeb\@extra@b@citeb}%
		\NAT@date
	}%
}
{%
	\@citea
	\NAT@nmfmt{\NAT@nm}%
	\NAT@aysep\NAT@spacechar
	\NAT@hyper@{\NAT@date}%
}
{}{}
\xpatchcmd\NAT@citex
{%
	\@citea\NAT@hyper@{%
		\NAT@nmfmt{\NAT@nm}%
		\hyper@natlinkbreak{\NAT@spacechar\NAT@@open\if*#1*\else#1\NAT@spacechar\fi}%
		{\@citeb\@extra@b@citeb}%
		\NAT@date
	}%
}
{
	\@citea
	\NAT@nmfmt{\NAT@nm}%
	\NAT@spacechar\NAT@@open\if*#1*\else#1\NAT@spacechar\fi
	\NAT@hyper@{\NAT@date}%
}
{}{}
\makeatother

\makeatletter
\renewcommand*\aa@pageof{, page \thepage{} of \pageref*{LastPage}}
\makeatother

\begin{document} 

   \title{Magnetic dynamos in galaxy clusters: the crucial role of galaxy formation physics at high redshifts}
   \titlerunning{Magnetic dynamos in galaxy clusters}
   \authorrunning{Tevlin et al.}

   \author{L. Tevlin 
          \inst{1,2}
          \and
          T. Berlok \inst{3,1}
          \and C. Pfrommer \inst{1}
          \and R. Y. Talbot \inst{4}
          \and J. Whittingham \inst{1,2}
          \and E. Puchwein \inst{1}
          \and R. Pakmor \inst{4}
          \and R. Weinberger \inst{1}
          \and V. Springel \inst{4}
          }

   \institute{Leibniz-Institute for Astrophysics Potsdam (AIP), An der Sternwarte 16, 14482 Potsdam, Germany
    \and
    Institut für Physik und Astronomie, Universität Potsdam, Karl-Liebknecht-Str. 24/25, 14476 Golm, Germany
    \and 
    Niels Bohr Institute, Blegdamsvej 17, DK-2100 Copenhagen, Denmark
    \and 
    Max Planck Institute for Astrophysics, Karl-Schwarzschild-Strasse 1, 85740 Garching, Germany
             }

   \date{Received \today}

  \abstract{
   Observations of Faraday rotation and synchrotron emission in galaxy clusters imply large-scale magnetic fields with $\upmu\mathrm{G}$ strengths possibly extending back to $z=4$. Non-radiative cosmological simulations of galaxy clusters show a comparably slow magnetic field growth that only saturates at late times. We investigate the effects of including galaxy formation physics and find a significantly accelerated magnetic field growth. After adiabatically compressing the magnetic seed fields, we observe further amplification by a fluctuation dynamo until reaching approximate energy equipartition with the turbulent flow. We identify three crucial stages in the magnetic field evolution. 1) At high redshift, the central dominant galaxy serves as the prime agent that magnetizes not only its immediate vicinity but also most of the forming proto-cluster through a combination of a small-scale dynamo induced by gravitationally-driven, compressive turbulence and stellar and active galactic nuclei feedback that distribute the magnetic field via outflows. 2) This process continues as other galaxies merge into the forming cluster in subsequent epochs, thereby transporting their previously amplified magnetic field to the intracluster medium through ram pressure stripping and galactic winds. 3) At lower redshift, gas accretion and frequent cluster mergers trigger additional small-scale dynamo processes, thereby preventing the decay of the magnetic field and fostering the increase of the magnetic coherence scale. We show that the magnetic field observed today in the \emph{weakly collisional} intracluster medium (ICM) is consistently amplified on \emph{collisional} scales. Initially, this occurs in the collisional interstellar medium during proto-cluster assembly, and later in the ICM on the magnetic coherence scale, which always exceeds the particle mean free path, supporting the use of magneto-hydrodynamics for studying the cluster dynamo. We generate synthetic Faraday rotation measure observations of proto-clusters, highlighting the potential for studying magnetic field growth during the onset of cluster formation at cosmic dawn.
}


   \maketitle
%

\section{Introduction}

Astrophysical magnetic fields can be found across a huge range of different scales, reaching from stars at the small scales to the intracluster medium (ICM) in galaxy clusters ($\sim \mathrm{Mpc}$). These magnetic fields can be inferred from Faraday rotation measures \citep[RMs,][]{2002ARA&A..40..319C, 2022A&A...665A..71O} and radio synchrotron observations \citep{2019SSRv..215...16V}. The discovery of radio megahalos even provides evidence for volume-filling magnetic fields out to the virial boundaries of galaxy clusters \citep{2022Natur.609..911C,2022SciA....8.7623B}. These observations reveal that galaxy clusters are filled with magnetic fields with strengths of a few $\upmu \mathrm{G}$ and coherence lengths at the order of a few $\mathrm{kpc}$ \citep{2005A&A...434...67V,2011A&A...529A..13K}. However, the generation and amplification mechanisms of magnetic fields within galaxy clusters remain uncertain.

Understanding cluster magnetic fields is crucial for understanding the diverse and rich radio phenomena in galaxy clusters that are linked to the transport and cooling of cosmic rays along magnetic field lines \citep{2023A&ARv..31....4R}. Yet, the exact origins of these phenomena such as radio halos, relics, diffuse radio emission near cluster centers, which reveal interactions between active galactic nuclei (AGN) feedback and the magnetized ICM, remain uncertain \citep{2014IJMPD..2330007B, 2019SSRv..215...16V, 2023MNRAS.523..701W}. Recent radio observations, conducted by the LOFAR Two-metre Sky Survey (LoTSS), have unveiled that galaxy clusters at redshifts around $z \sim 1$ have magnetic field strengths similar to low redshift clusters \citep{2021NatAs...5..268D, 2023A&A...675A..51D}. This implies an early phase of magnetic field creation and amplification within galaxy clusters. Observations of high-redshift protoclusters in the redshift range $2.2 < z < 4.25$ have unveiled interactions between the brightest cluster galaxy (BCG) and the magnetized ICM \citep{2022ApJ...937...45A, 2023A&A...676A..29C, 2024ApJ...961..120C}, which gives hints that they might play a role in the high redshift magnetization of the protocluster. 

There is consensus that magnetic field growth in galaxy clusters starts with a small seed field that gets amplified. The seed field's origin can be attributed to magnetic field generation in the early Universe, such as through the Biermann battery mechanism that turns baroclinic motions in ionized gas into magnetic fields \citep{1950ZNatA...5...65B, 1997ApJ...480..481K, 2022hxga.book...56K}, the Weibel instability \citep{1959PhRvL...2...83W, 2015NatPh..11..173H, 2024ApJ...960...12Z}, or cosmological phase transitions in the early Universe \citep{2002RvMP...74..775W,2013A&ARv..21...62D}. The most widely discussed amplification scenario involves the turbulent small-scale dynamo, which converts turbulent kinetic energy into magnetic energy \citep{1968JETP...26.1031K, 1966ITAP...14..389K, 2019JPlPh..85d2001R, 2022hxga.book...56K}. Dynamo theory makes use of the fact that a turbulent magnetic field, flux-frozen into the thermal plasma, can evolve below the viscous scales, where the turbulence in the velocity field is dissipated, and grows rapidly at the resistive scales where timescales are very short (see e.g., \citealt{2022hxga.book...56K}). Turbulence within the ICM arises mostly from mergers and interactions with galaxies \citep{1997ApJ...480..481K, 1999LNP...530..106N, 2006MNRAS.366.1437S,10.1093/mnras/sts439, 2017MNRAS.471.3212W, 2017MNRAS.464..210V, 2020ApJ...889L...1L, 2023ApJ...950...24B}. It can also result from plasma instabilities such as the magneto-thermal instability \citep{2000ApJ...534..420B,2024A&A...682A.125P,2024arXiv240206718P}, pressure anisotropies \citep{2024A&A...683A..35R}, and magnetic pumping
\citep{2024ApJ...965..155L} that may all be able to self-regulate magnetic field growth. 

According to the theory of hierarchical structure formation, galaxy clusters form through frequent mergers with galaxies and smaller clusters over cosmic time \citep{2012ARA&A..50..353K}. In this paper, we demonstrate how the magnetic evolution of galaxy clusters is linked to that of the galaxies themselves. During the collapse of the protocluster, the primordial magnetic field is adiabatically compressed and turbulence is generated, which initiates dynamo processes as seen in cosmological simulations of galaxy clusters (e.g., \citealt{2014MNRAS.445.3706V, 2019MNRAS.486..623D, 2024ApJ...967..125S}). In galaxies, gravitational infall and stellar feedback processes drive turbulence, which further contributes to dynamo processes (e.g., \citealt{2013A&A...560A..87S, 2018A&A...611A..15K, 2022MNRAS.515.4229P}). After a seed magnetic field has been amplified through a small-scale dynamo in galaxies, it can be transported and expanded into intergalactic space by powerful galactic winds or jets, or ram pressure stripping   \citep{Rees1968,1987QJRAS..28..197R,1988ComAp..12..265R,2001ApJ...556..619F,2004PASJ...56...29F, 2014MNRAS.442..406H, 2016IAUFM..29B.699D, 2019ApJ...884L..45M, 2023A&A...673A.146S}. These mechanisms impact galaxies within the cluster, but also in pre-merging substructures, a phenomenon known as ``pre-processing'' \citep{2004PASJ...56...29F, 2023A&A...673A.146S} in the context of metal enrichment. Observations and simulations support this, showing that many galaxies in cluster environments lose their self-bound gas reservoirs \citep{1984ApJ...285..426B, 2014MNRAS.442..406H, 2016IAUFM..29B.699D, 2019ApJ...884L..45M, 2023ApJ...950..114H}. Clusterrs additionally inject turbulence into the ICM, which enables further dynamo activity (e.g., \citealt{2014MNRAS.445.3706V, 2019MNRAS.486..623D, 2024ApJ...967..125S}).
   
Numerical simulations serve as valuable tools for investigating the evolution of magnetic fields in galaxy clusters (see e.g. the review by \citealt{Donnert2018}). Previous studies have explored the creation and amplification of magnetic fields in isolated galaxy clusters \citep[without radiative physics,][]{2019ApJ...883..138R}, galaxy clusters in cosmological simulations \citep[without radiative physics,][]{1999A&A...348..351D, 2015Natur.523...59M, 2018MNRAS.474.1672V, 2022ApJ...933..131S, 2024ApJ...967..125S}, and cosmological simulations with passive magnetic fields \citep{2020BAAA...61Q..80A} or active magnetic fields while including radiative physics and subgrid models for galaxy formation \citep{2015MNRAS.453.3999M,Marinacci2018,Nelson2024}. Most simulations successfully reproduce the magnetic field evolution to match observed values of a few $\upmu \mathrm{G}$ within the age of the universe. 

Such simulations necessitate the coverage of a wide range of scales, spanning from small events like supernovae (SN) to large events such as cluster mergers. In ideal magneto-hydrodynamical (MHD) simulations, the growth rates depend on the cell sizes, as these determine the shortest eddy turnover scales \citep[see e.g.,][]{2017MNRAS.469.3185P,2021MNRAS.506..229W,2022MNRAS.515.4229P}. Therefore, simulations require high resolution to reproduce the strong magnetic fields observed at high redshifts. While \cite{2022ApJ...933..131S} found that the magnetic field only recently became saturated, the higher resolved simulations of \cite{2024ApJ...967..125S} observed a saturated magnetic field already at  $z\approx4$.

However, increasing the resolution beyond the viscous scales theoretically necessitates the inclusion of kinetic plasma effects (e.g. \citealt{Berlok2020,2020JPlPh..86e9003S,2024ApJ...967..125S}). A related question is whether dynamo theory is applicable in the ICM, given that Kazantsev theory is a fluid theory. Due to the weakly collisional nature of the ICM (see e.g., \citealt{2006PhPl...13e6501S}, \citealt{2011MNRAS.410.2446K}, \citealt{2012ApJ...754..122K} and references therein), the fluid approximation is formally invalid, and thus any MHD description of the ICM would also be invalid. Additionally, microscale kinetic instabilities, specifically the firehose and mirror instabilities on small scales (see e.g., \citealt{2006PhPl...13e6501S} and \citealt{2024arXiv240502418M}), can increase the effective collision rate and reduce the effective viscosity. Cluster observations indicate that the turbulent dissipation scale is smaller than expected from Spitzer viscosity \citep{Zhuravleva_2019,Heinrich2024}, which is theorized to be due to these small scale instabilities. However, in high-beta plasmas such as the ICM \citep{2002ARA&A..40..319C}, it has been argued that the pressure anisotropy will self-regulate by reorganizing turbulent motions \citep{2019JPlPh..85a9014S, 2024arXiv240502418M}, and that this so-called magneto-immutability \citep{2019JPlPh..85a9014S} can reduce the viscosity without triggering the small scale instabilities \citep{2024arXiv240502418M}. Simulations that include kinetic plasma instabilities via MHD with anisotropic viscous stress \citep{1965RvPP....1..205B} show only minor deviations from the saturated state of the normal MHD dynamo \citep{2018ApJ...863L..25S, 2019JPlPh..85a9014S,2020JPlPh..86e9003S} and hybrid-kinetic particle-in-cell simulations show a similar magnetic field growth to the MHD dynamo model \citep{2024MNRAS.528..937A}. 

However, these studies only considered magnetic dynamo action in weakly-collisional conditions that are found in the ICM at the current epoch and have not self-consistently accounted for the magnetic dynamo in the plasma that evolves into the ICM throughout cosmic evolution. MHD simulations that include radiative physics can naturally resolve higher growth rates due to higher densities and smaller cell sizes in the interstellar medium (ISM). In the ISM, the fluid description is valid again due to the higher particle collision rates. It has been shown that simulations of magnetic fields in galaxies reveal rapid early growth \citep{2008A&A...486L..35G, 2010A&A...522A.115S, 2014ApJ...783L..20P, 2017MNRAS.469.3185P}. This fast growth is linked to supernovae from early generations of stars and gravitational collapse due to structure formation processes \citep{2013A&A...560A..87S, 2018A&A...611A..15K, 2022MNRAS.515.4229P}. Cosmological galaxy simulations have also shown that magnetized galactic outflows can occur and magnetize their surroundings \citep{2020MNRAS.498.3125P, 2021MNRAS.501.4888V}. This provides a method to magnetize the ICM in radiative simulations: magnetic fields can grow rapidly in the ISM, and the magnetized gas is then transported outward via feedback mechanisms.

\cite{2015MNRAS.453.3999M} demonstrated that incorporating radiative physics results in faster magnetic field growth in the ICM compared to non-radiative simulations, although they did not investigate the exact sources of this accelerated growth. This leaves the the contributions of galaxies to the magnetic field evolution in galaxy clusters unclear. In this paper, we expand on the work of \cite{2015MNRAS.453.3999M} by comparing radiative and non-radiative simulations of galaxy clusters. We employ higher resolution and updated galaxy formation physics. Furthermore, we extend the dynamo analysis to shed light on the mechanisms of magnetic field growth and transport.

In this study, we conduct two sets of each four cosmological galaxy cluster simulations, one set with radiative physics enabled and the other set with radiative physics disabled. We show that the magnetic field grows faster in the radiative simulations due to the enhanced growth rate inside galaxies. Furthermore, we show how the magnetic field in the clusters grows with subsequent mergers of galaxies. The paper is divided as follows. In Sect.~\ref{sec:methods}, we describe our simulation setup. We give properties of the simulated clusters, introduce the cosmological code and model, and the galaxy formation model. In Sect.~\ref{sec:3}, we show how the magnetic field evolves with time in the two simulation sets. In Sect.~\ref{sec:4}, we investigate the origin of the magnetic field growth, including a detailed analysis on the dynamo activity in our simulations. In Sect.~\ref{sec:6}, we present mock high redshift Faraday RM observations of the protocluster. Finally, in Sect.~\ref{sec:discussion}, we discuss our findings, summarize our most important results, and give an outlook.

\section{Methods}
\label{sec:methods}

We analyze the magnetic field evolution in four different galaxy clusters which have been simulated with two distinct physical models using the zoom-in technique. One type of simulation uses MHD and gravity (`non-radiative' simulations), while the other one additionally includes subgrid models for galaxy formation and feedback physics (`radiative' simulations). These simulations are part of the PICO-Clusters simulation suite, which will be described in detail in a forthcoming paper (Berlok et al, in prep.). The PICO-Clusters simulations employ the moving-mesh code AREPO \citep{Springel_2010, 2016MNRAS.455.1134P, 2020ApJS..248...32W} for evolving the MHD equations \citep{2013MNRAS.432..176P} in comoving form and uses an updated version called AREPO-2 that includes the SUBFIND-HBT algorithm of GADGET4 \citep{2021MNRAS.506.2871S, 2022ascl.soft04014S}. To simulate radiative physics, the galaxy formation model used in the IllustrisTNG simulations \citep{2017MNRAS.465.3291W,2018MNRAS.473.4077P} is employed. Brief descriptions of AREPO and its ability to correctly evolve magnetic fields (Sect.~\ref{sec:arepo-mhd}), IllustrisTNG (Sect.~\ref{sec:illustris-tng}) and PICO-Clusters (Sect.~\ref{sec:pico}) are provided below. We summarize the details most important to this paper in order to keep it self-contained. An overview of the physical characteristics of the four selected clusters at redshift zero is shown in Table~\ref{table:1}, including the virial radius, virial mass, and the number of subhalos at $z=0$. Here the virial radius ($R_{200}$) is the radius that contains a mean density of 200 times the critical density, and the virial mass ($M_{200}$) is the mass contained in a sphere of this virial radius. We also show the number of high resolution dark matter particles in each zoom12 simulation (see Sect.~\ref{sec:pico}), each of which have a
mass of $5.9 \times 10^7~\mathrm{M}_\odot$. This makes the mass resolution in our simulations comparable\footnote{While we adopted the same particle masses as in TNG300, we use a somewhat different cosmology, which effectively implies a different effective mass resolution.} to the TNG300 cosmological volume simulation \citep{2018MNRAS.475..676S} and the TNG-Cluster zoom-in simulations \citep{Nelson2024}. The halo numbers are assigned by sorting the $M_{200}$ masses of all clusters with $M_{200} > 10^{15}~\mathrm{M}_\odot$ at $z=0$ in the parent simulation, starting with the most massive cluster, Halo0. As a result, ``Halo3'' represents the fourth most massive halo in our simulation.

We chose to use a sample of four clusters to test for generality of the suggested mechanism of the cluster magnetic dynamo and limit the impact of selection effects. We then perform a more in-depth analysis of one of the clusters (Halo3) in order to elucidate the detailed physics of the mechanism of magnetic field growth.
Halo3 and 4 are two of the most massive clusters in the PICO-Clusters sample and were selected for re-simulation because of their massive, low redshift, galaxy cluster mergers. Halo194 and 260 were randomly selected among the PICO-Clusters that have $M_{200} \approx 10^{15} \mathrm{M}_\odot$ and were included in the analysis in this paper in order to ensure that our results are applicable to somewhat smaller clusters and not just to very massive clusters undergoing major mergers.

In our simulation analysis, we will distinguish between ICM and ISM by designating star-forming cells as ISM and non-star-forming cells within the cluster virial radius as ICM. When we refer to the ``center'' of the galaxy cluster, this is to be specifically understood as a spherical region of cluster-centric physical radius $20~\mathrm{kpc}$, which corresponds to the virial radius of the central galaxy in Halo3 at $z=9.5$. We make this separation to distinguish between actions occurring in the BCG at high redshift and activity in the entire cluster,\footnote{For simplicity we refer to the central galaxy and halo of the main cluster progenitor by BCG and cluster, respectively, even if these terms may not be fully appropriate at high redshift due to the smaller masses of the objects.} both of which contribute to the magnetic field enrichment.

\begin{table}
\caption{ID, size, mass, number of subhalos and resolution of selected halos in the radiative zoom12 simulations at $z=0$.}
\label{table:1}
\centering
\begin{tabular}{c r c c c c c}
\hline\hline
Halo & FoF ID & $R_{200}$ & $M_{200}$ & $\mathrm{N}_\mathrm{subhalos}$  & $N_\mathrm{DM}$ \\
 &  & $[\mathrm{Mpc}]$ & $[10^{15} \mathrm{M}_\odot]$ & & \\
\hline
   3 & 11 &2.96 & 2.72 & 7098 &   $1.3 \times 10^8$\\
   4 & 6 &2.96 & 2.72    & 9109 &  $1.3 \times 10^8$ \\
   194 & 374 &2.15 & 1.04     & 3972  & $0.6 \times 10^8$ \\
   260 & 321 &2.22 & 1.16    & 4743  & $0.5 \times 10^8$\\
\hline
\end{tabular}
\end{table}

\subsection{AREPO and magnetic fields}
\label{sec:arepo-mhd}
AREPO discretizes MHD quantities on a moving unstructured mesh that is defined by the Voronoi tessellation of a set of discrete points that can be moved arbitrarily. If those mesh-generating points follow the motion of the gas, the code inherits the advantages of Lagrangian methods. The Voronoi mesh can adaptively refine and de-refine. The standard refinement criteria in AREPO ensures that cell masses remain within a factor of 2 of some target mass, but additional criteria can be used to enhance the resolution in areas of interest relevant to the particular simulation in question. Dark matter, stars, and black holes are included as collisionless particles. Gravitational interactions are computed using a TreePM method \citep{2005MNRAS.364.1105S} to separate long- and short-range gravitational forces. Here the long-range forces are determined using a Fourier transform on a mesh, while the short-range forces are solved using an oct-tree algorithm \citep{1986Natur.324..446B}.

The comoving ideal MHD equations are solved in AREPO on the moving Voronoi mesh \citep{2011MNRAS.418.1392P,2013MNRAS.432..176P}. The equations are solved using an HLLD Riemann solver \citep{2005JCoPh.208..315M} and the code is able to reproduce comoving analytic wave solutions for both Alfvén and magnetosonic waves \citep{2022MNRAS.515.3492B}. To mitigate divergence errors in the magnetic field, the Powell 8 wave divergence control scheme \citep{1999JCoPh.154..284P} is utilized, which subtracts the divergence error from the initial state, enhancing overall accuracy and ensuring stability. Simulations that use this MHD scheme can reproduce analytic growth rates and magnetic curvature statistics expected for a fluctuating dynamo \citep{2022MNRAS.515.4229P} as well as different observables, e.g., magnetic field saturation strengths and radial profiles in disk galaxies \citep{2014ApJ...783L..20P, 2017MNRAS.469.3185P}, Faraday RMs in spiral galaxies and the Milky Way \citep{2018MNRAS.481.4410P,2023NatAs...7.1295R}, and RM values of fast radio burst in normal star forming galaxies \citep{2023ApJ...954..179M}. Furthermore, the MHD setup is able to reproduce observed magnetic fields in dwarf galaxies, galaxies and groups \citep{2024MNRAS.528.2308P}, and in the circumgalactic medium \citep{2020MNRAS.498.3125P} via dynamo activity. 
The divergence error due to the Powell cleaning method remains low in dynamic galaxy mergers \citep{2021MNRAS.506..229W, 2023MNRAS.526..224W}, which shows that magnetic field amplification is not an artifact of the code. We adopt an initially homogeneous magnetic seed field and initialize a comoving magnetic seed field of $10^{-14}~\mathrm{G}$ at $z=127$. Variations in seed strength show little to no impact on the later evolution as the initial memory is quickly erased by dynamo activity \citep{2013MNRAS.432..176P, 2014ApJ...783L..20P, 2015MNRAS.453.3999M}.

The efficiency of the fluctuating dynamo, as well as the morphology of the resulting magnetic field, depend on the smallest scales, particularly the viscous and resistive scales. While the ideal MHD equations assume a perfectly conducting fluid (neglecting resistivity and viscosity), numerical effects stemming from finite volume and time discretization introduce numerical viscosity and resistivity. The numerical resolution defines the numerical viscous and resistive scales, equivalent to the smallest eddy scales. The utilized (quasi-)Lagrangian moving mesh with an approximately constant gas mass per resolution element yields higher resolution in dense regions (with cell sizes of $\sim 100~\mathrm{pc}$ in the center) and degrades the resolution in low-density regions (reaching resolutions of $\sim 10-100~\mathrm{kpc}$ in the outskirts). This enables us to resolve resolution-dependent Reynolds numbers of $\sim 100-1000$ in the ICM.

\subsection{The IllustrisTNG galaxy formation model}
\label{sec:illustris-tng}

We simulate radiative physics using the IllustrisTNG galaxy formation model \citep{2017MNRAS.465.3291W,2018MNRAS.473.4077P}, which encompasses star formation and feedback \citep{2013MNRAS.436.3031V, 2018MNRAS.473.4077P}, supermassive black hole seeding, growth, and resulting AGN feedback \citep{2005ApJ...620L..79S, 2005MNRAS.361..776S,2013MNRAS.436.3031V, 2017MNRAS.465.3291W}, as well as gas cooling and heating. Simulations using the IllustrisTNG model have been shown to reproduce observed stellar-halo mass functions of groups and galaxy clusters \citep{2018MNRAS.473.4077P}, and clustering properties, such as two-point correlation functions between low-redshift galaxies that match measurements from SDSS data \citep{2018MNRAS.475..676S}. Magnetic field strengths of galaxy clusters, simulated using this model, are consistent with observations \citep{2018MNRAS.480.5113M}. In this paper, we expand on the work of \citet{2015MNRAS.453.3999M}, who used the Illustris galaxy formation model \citep{2014MNRAS.444.1518V} and additionally account for the magnetic field evolution using ideal MHD.

To model star formation and the ISM, we employ an effective model \citep{2003MNRAS.339..289S, 2013MNRAS.436.3031V, 2018MNRAS.473.4077P}. The formation of star particles occurs stochastically when the density exceeds a threshold of $n_\mathrm{H} > 0.106~\mathrm{cm}^{-3}$ adhering to the Kennicutt-Schmidt relation that is based on observations \citep{1959ApJ...129..243S, 1989ApJ...344..685K}, and using a Chabrier initial mass function \citep{2003PASP..115..763C}. Mass loss and metal enrichment are calculated in each time step, considering the initial mass function, as outlined in \cite{2018MNRAS.473.4077P}. Stellar feedback employs a kinetic wind model, wherein wind particles are decoupled from hydrodynamics, interacting only gravitationally. They recouple hydrodynamically upon encountering gas cells that fulfill a density threshold ($n_\mathrm{H} < 0.0053~\mathrm{cm}^{-3}$, which corresponds to 5 percent of the star formation threshold) or, alternatively, after a defined travel time $0.025 \times 1/H(z)$, where $1/H(z)$ is the Hubble time at the simulation redshift. At very high redshift, recoupling is primarily governed by the time criterion due to shorter Hubble times and higher critical densities. Upon recoupling, wind particles inject mass, metals, thermal energy, and momentum to gas cells. Wind particles form stochastically from star forming cells and are launched isotropically, with velocity correlated to local dark matter velocity dispersion and adjusted for redshift.

We include gas cooling and heating following the method outlined in \citet{2013MNRAS.436.3031V}. A treatment for primordial cooling and heating through ionization equations based on cooling, recombination, and ionization rates for primordial gas is employed \citep{1992ApJS...78..341C, 1996ApJS..105...19K}, alongside Compton cooling off of the cosmic microwave background \citep{1986ApJ...301..522I}. Metal line cooling uses CLOUDY cooling tables \citep{2009MNRAS.393...99W}. A spatially uniform, redshift-dependent ionizing UV background is included for gas heating \citep{2009ApJ...703.1416F} with corrections for self-shielding in dense gas \citep{2013MNRAS.430.2427R}.

The AGN treatment in IllustrisTNG is based on the AGN model developed by \citet{2005MNRAS.361..776S} and further developed by \citet{2013MNRAS.436.3031V} and \citet{2017MNRAS.465.3291W}. Each halo that reaches a friends-of-friends halo mass of $5 \times 10^{10}~\mathrm{M}_\odot$ is assigned a black hole at the position of the densest cell in the halo \citep{2017MNRAS.465.3291W}. The black holes in the simulation accrete gas from their surrounding environment. A fraction of the accreted gas contributes the black hole mass. A fraction of the rest mass energy of the accreted gas is expelled as energetic feedback, some of which couples to the surrounding gas. The feedback model employed in this study is a two-state model, where the feedback mode depends on the accretion rate. This differentiation is based on the mass accretion rate in units of the Eddington rate ($\dot{M}_{\mathrm{BH}}/\dot{M}_{\mathrm{Edd}}$) that fulfills a threshold value that scales with black hole mass, but should not exceed 0.1. Below this threshold, the black hole operates in a kinetic AGN feedback mode, generating black hole-driven winds \citep{2017MNRAS.465.3291W}. Otherwise, the black hole operates in quasar mode, releasing thermal energy isotropically to neighboring gas cells \citep{2005ApJ...620L..79S, 2005MNRAS.361..776S}. The quasar mode is predominantly active at higher redshifts.

\subsection{PICO-Clusters and halo selection}
\label{sec:pico}

PICO-Clusters (Plasmas In COsmological Clusters, Berlok et al., in prep.) is a new simulation suite of cosmological zoom-in simulations of galaxy clusters using the IllustrisTNG galaxy formation model with the moving mesh code AREPO-2 (an updated version of AREPO that evolved from the MilleniumTNG project \citep{2023MNRAS.524.2539P}; for details, see below). It consists of a cosmological volume simulation of a periodic box with side length $\approx 1.5$ Gpc that contains 272 galaxy clusters with masses above $10^{15}~\mathrm{M}_\odot$
and re-simulations of 25 clusters using the zoom-in technique. The long-term goal of PICO-Clusters is to understand the role of including various (plasma-)physical models in simulations, such as anisotropic viscosity \citep{Berlok2020}, anisotropic conduction \citep{Talbot2024}, cosmic ray protons \citep{2016PakmorCRs, Pfrommer2017} and electrons \citep{Winner2019}, and AGN feedback via a low-density momentum-driven jet \citep{Weinberger2017,Ehlert2023}.
The base-line physical model of PICO-Clusters is the IllustrisTNG galaxy formation model (see Sect.~\ref{sec:illustris-tng}). This makes
the 25 base-line PICO-Clusters simulations roughly comparable to the TNG300 and TNG-Cluster simulations \citep{2018MNRAS.473.4077P,Nelson2024}. In the present paper, we start this upcoming systematic investigation of the role of the physical model employed by comparing non-radiative simulations with the base-line IllustrisTNG model with a focus on magnetic fields.

All PICO-Cluster simulations are initialized at $z=127$ and have simulation properties stored in snapshots. The four clusters analyzed in this paper have a high output frequency, i.e., around 260 full snapshots per simulation. This allows for a detailed analysis of the time evolution. In addition, the AREPO-2 code has better memory management and load balancing, and includes the SUBFIND-HBT algorithm \citep{2021MNRAS.506.2871S, 2022ascl.soft04014S} that allows storing group and subfind properties even more frequently than the full snapshots and to easily calculate merger trees. We use the on the fly merger trees for tracking the centers of the galaxy clusters \citep{2022ascl.soft04014S}.

PICO-Clusters uses a Planck-2018 cosmology \citep{PlanckCollaboration2020} with the following cosmological parameters $\Omega_{\Lambda} = 0.684$, $\Omega_{\mathrm{m}0} = 0.316$, $\Omega_{\mathrm{b}0} = 0.049$ and the Hubble parameter $h=0.673$. We used these parameters to create initial conditions for the parent simulation. We generated a power spectrum using the CAMB code \citep{Lewis:1999bs, Lewis:2002ah}, and N-GenIC \citep{Springel2015} was then used to create initial conditions based on Gaussian random fields, adhering to the power spectrum prescription (as detailed in \citealt{2005Natur.435..629S, 2012MNRAS.426.2046A}). The parent simulation had $1024^3$ dark matter particles, each with a mass of $1.0 \times 10^{11} \mathrm{M}_\odot$. This cosmological box was evolved to $z=0$. The most massive halos at $z=0$ were identified using a friends-of-friends algorithm (FoF). The initial conditions for the re-simulations were created using a new code that will be described in detail in Puchwein et al, in prep. In brief, the initial conditions are set up such that each dark matter particle has a mass of $5.9 \times 10^7~\mathrm{M}_\odot$ inside a high-resolution region (for reference, this corresponds to the TNG300 resolution). The results presented in this paper are based on the zoom12 simulations, where the zoom-in regions have a mass resolution that is $12^3$ times higher than that of the parent simulation. To ensure a smooth transition between the high-resolution region inside the halo and the lower resolution outside, a third dark matter particle type with intermediate mass was introduced. The end result is that the clusters analyzed in this paper have no low resolution particles inside $3R_{200}$ at $z=0$. The target gas mass is $7 \times 10^6~\mathrm{M}_\odot$. The high resolution dark matter, star and black hole particles have comoving softening lengths of $4.8~\mathrm{kpc}$ above redshift 1 and a physical softening of $2.4~\mathrm{kpc}$ at lower redshift. The gas softening depends on the cell size with a minimum comoving value of $0.6~\mathrm{kpc}$.

\section{Magnetic field evolution with time}
\label{sec:3}

In this section, we will compare magnetic field growth in the radiative and non-radiative simulations and explore environmental effects for the differences such as interactions between the ICM and galaxies. 

\subsection{Projections of magnetic field and vorticity}

\begin{figure*}
\centering\includegraphics[width=1.0\textwidth]{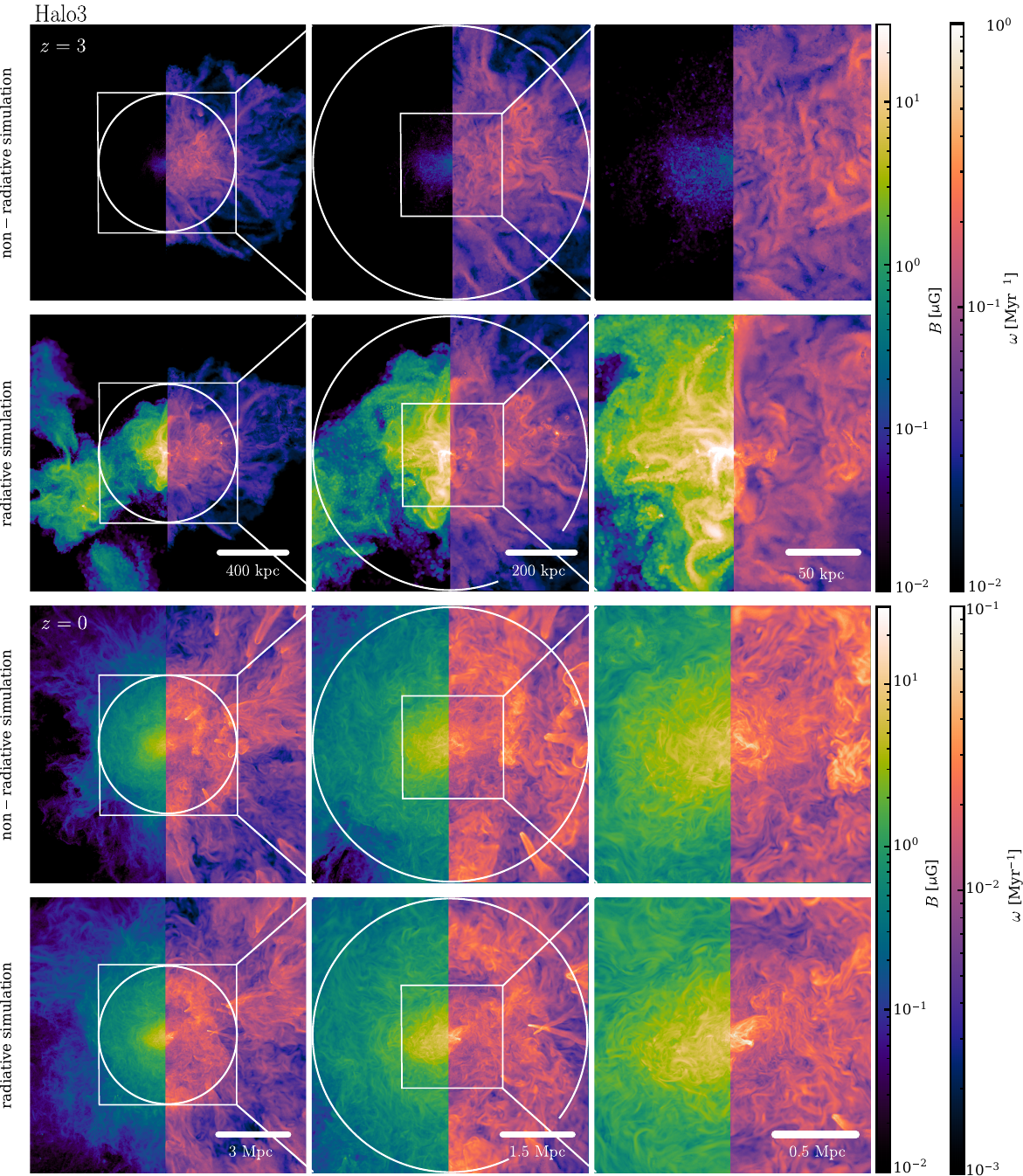}
  \caption{Thin projections through the center of Halo3 displaying the root mean square magnetic field strength and vorticity, respectively, at $z=3$ in the non-radiative (first row) and radiative (second row) simulations. The bottom two rows show the same but at $z=0$. The virial radius $R_{200}$ is indicated with a white circle. From left to right, the panels have sidelengths $L=4 R_{200},~2 R_{200},~0.5 R_{200}$ and a thickness of $L/10$. At $z=3$, the magnetic field in the radiative simulation is nearly saturated. The BCG appears as a bright spot with a strong magnetic field. An object infalling from the left has also developed a strong magnetic field that will contribute to the cluster's final magnetic field. By contrast, at $z=3$ the magnetic field in the non-radiative simulation remains weak. The infalling object is not visible, and there is no BCG. This underscores the magnetic field growth in galaxies and its transport to the ICM via merging halos.
  }
  \label{fig:projection_z3}
\end{figure*}

We present in Fig.~\ref{fig:projection_z3} thin projections of Halo3 at $z=3$ (upper panels) and at $z=0$ (lower panels) in both simulation setups. The virial radius ($R_{200} = 380~\mathrm{kpc}$ at $z=3$ and $R_{200} = 3~\mathrm{Mpc}$ at $z=0$) is indicated with a white circle. From left to right, we subsequently zoom into the same projection. In each panel we show the magnetic field strength (left) and vorticity (right). At $z=3$, the magnetic field is significantly larger in the radiative simulation whereas it looks very similar between the two simulations setups at $z=0$.

At $z=3$ (upper panel of Fig.~\ref{fig:projection_z3}), a strong magnetic field with $30~\upmu \mathrm{G}$ has developed in the center of the radiative simulation. The magnetic field fills the volume in the cluster, with field strengths decreasing towards the outskirts to $\sim 10^{-2}~\upmu \mathrm{G}$ in the radiative simulation. The location of the BCG in the center shows a bright spot of magnetic field and reflects the very efficient amplification in the central galaxy. A merging cluster is visible on the left side. The merging cluster has grown its own strong magnetic field that will contribute to the main halo's magnetic field after merging. Both findings highlight the physical mechanisms at work to amplify the magnetic field: the growth in galaxies and enrichment of the ICM via subsequent mergers. By comparison, in the non-radiative simulation, the magnetic field remains weak and only grows to $\sim 10^{-2}~\upmu \mathrm{G}$ in the center, where the densities are highest. The merging cluster has not grown a magnetic field that is above the plotting threshold of $10^{-2}~\upmu \mathrm{G}$.

Turbulence and magnetic field strength in the ICM are intricately linked through potential dynamo interactions. Projections of the root mean square vorticity, which itself is defined as $\bm\omega = \bnabla \btimes \bm{\varv}$, look comparable in both simulations at $z=3$. Remarkably, central filaments with the highest vorticity also coincide with regions displaying the strongest magnetic fields in the radiative simulation. Eddy turnover rates in the central regions are approximately $1 ~\mathrm{Myrs}^{-1}$ in both simulations, indicating a similar vorticity profile within and beyond the virial radius in the two simulations. This turbulence is primarily attributed to frequent mergers and accretion shocks in both simulations, rather than originating from radiative physics, which would naturally be only visible in the radiative simulation. Various studies come to the same conclusions \citep{1999LNP...530..106N, 2011A&A...529A..17V, 2017MNRAS.464..210V}. Galaxies in the radiative simulation are visible as few bright spots, reflecting the enhanced level of turbulence injected by stellar and AGN feedback, as well as the small cell sizes.

In the lower panels, we compare the magnetic field strength and vorticity at $z=0$. Both look very similar in the two simulations. In both runs, the highest magnetic field values -- approximately $30~\upmu\mathrm{G}$ -- are observed in the central regions coinciding with the highest density regions. Magnetic field strength exponentially decreases towards the outskirts in both simulations. Beyond the virial radius, the magnetic field is slightly more extended in the radiative simulation. The highest vorticity values occur at the center, where density is greatest, resulting in smaller cell sizes and elevated vorticity. Jellyfish galaxies are visible as bright spots with tails that are moving towards the cluster center. These phenomena are also observed in low redshift galaxies within galaxy clusters and groups \citep[e.g.,][]{2017ApJ...844...48P} as well as in cosmological simulations \citep[e.g.,][]{2023MNRAS.524.3502R} and dedicated wind-tunnel simulations \citep[e.g.,][]{2012MNRAS.422.1609T,2024MNRAS.527.5829S} using similar galaxy formation physics.

The substantial vorticity levels in the ICM, and the strong spatial correlation between vorticity and magnetic field, is a first hint at the presence of a fluctuating dynamo mechanism in the ICM. Furthermore, this dynamo effectively amplifies magnetic fields to comparable strengths within the virial radius in both simulations at $z=0$. Subsequent sections will quantitatively analyze magnetic field evolution and identify the sources of magnetization.

\subsection{Radial magnetic field evolution in all four halos}

\begin{figure*}
  \includegraphics[width=\textwidth]{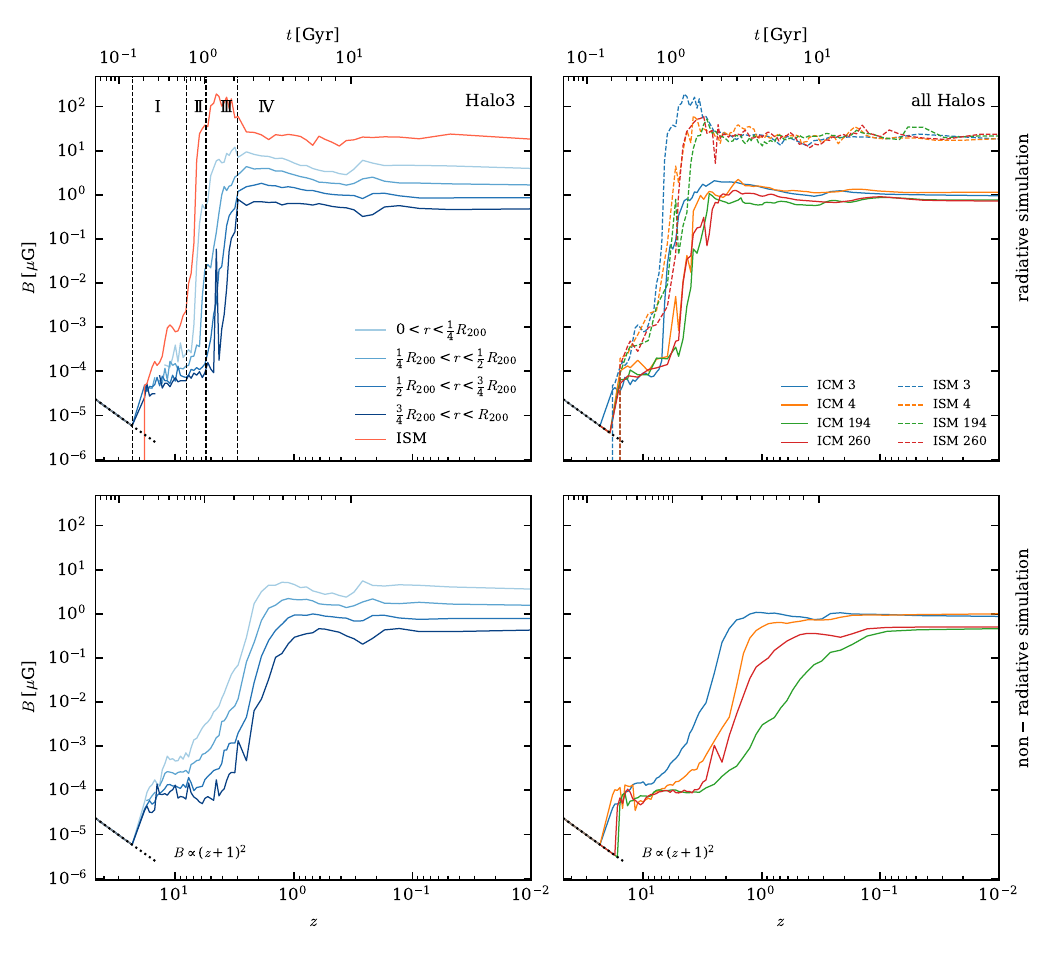}
  \caption{Evolution of magnetic field strength (derived from volume-weighted magnetic energy densities) in radiative (upper panels) and non-radiative (lower panels) simulations within $R_{200}$. The left panels focus on Halo3, the right panels show the magnetic field strength across four halos. The magnetic field amplification is faster in radiative simulations. Its evolution in the entire ICM closely mirrors that of the ISM across all 4 halos To show this in more detail, we mark 4 phases of magnetic field evolution in the radiative simulation of Halo3: \uproman{1}) the collapse of the first galaxies, \uproman{2}) the amplification in the BCG, \uproman{3}) the amplification in the host cluster by mergers that contribute their magnetically pre-enriched gas to the host halo, \uproman{4}) low redshift mergers that constantly inject turbulence, which drives a fluctuating small-scale dynamo.}
  \label{fig:B_time}
\end{figure*}

In order to \emph{i)} demonstrate that the magnetic field grows faster in the radiative simulations of all four halos, \emph{ii)} connect the growth in the ISM and the ICM, and \emph{iii)} distinguish between different phases of magnetic field growth, we analyze the time evolution of the magnetic field strength. Figure~\ref{fig:B_time} shows the time evolution of the volume-averaged root mean square magnetic field strength in the radiative simulation (upper panels), and in the non-radiative simulation (lower panels). In the left panels, we show as an example the evolution of the magnetic field in the ICM for Halo3 in more detail, including different radial ranges within the virial radius, indicated with the blue color scheme. The magnetic field in the ISM (i.e., by our definition, star-forming cells) is shown in red. 

In the right panels, we show the magnetic field in the ICM averaged across the virial radius for all four halos with solid lines. Dashed lines show the volume-averaged root mean square magnetic field strength in the ISM. The initial decay of the magnetic field strength at high redshift (shown in gray) reflects adiabatic expansion in an isotropically expanding Universe while conserving magnetic flux ($B \propto (1+z)^2$). We measure this in the entire high-resolution region. At around $z=20$ the protocluster starts to collapse and adiabatically compresses the field. The black dotted lines indicate the expected power-law fits to the adiabatic expansion before the protocluster collapse. 

Substantial differences between both simulation setups emerge during the high-redshift interval $20\gtrsim z>3$. We highlight four key stages in the magnetic field evolution within the radiative simulation of Halo3 (upper left panel).

\begin{enumerate}[I.]
\item $20 \gtrsim z > 8.5$: This phase marks the collapse of the first galaxies, which adiabatically compresses the initial seed magnetic field. Structure formation shocks introduce turbulence, which amplifies the magnetic field via a small-scale dynamo. This works very efficiently in the central, cool regions with short eddy turnover times.

\item $8.5>z>5.5$: At these high redshifts, the protocluster only consists of a few galaxies -- the progenitors that eventually merge to form the BCG. Supersonic cooling flows penetrate deep into the central regions and into star-forming galaxies, and inject turbulence. This leads to an exponential magnetic field growth in the cold ISM driven by the gravitational dynamo (we demonstrate this in Sect.~\ref{sec:4}). Stellar feedback produces additional turbulence, mainly within $R \lesssim 0.5 R_{200}$, and efficiently mixes the magnetically enriched gas in the galaxies with the ICM (shown in Sect.~\ref{ssec:winds}). As a result, the radiative simulation displays strong magnetic growth in the ISM, followed by the inner radial bin, and less pronounced growth in the outer radial bin.

\item $5.5>z>3$: This phase is marked by the first series of major mergers. Subsequent magnetic field enrichment results from mergers, contributing pre-enriched gas and turbulence to the ICM, leading to exponential growth (as shown in Sect.~\ref{sec:5}).

\item $3>z$: Left alone, the magnetic field would relax into a minimum energy configuration \citep{1974PhRvL..33.1139T}. The accretion of primordial (less) magnetized gas from the intergalactic medium would furthermore cause the magnetic field to decrease as a result of adiabatic cooling and mixing. Instead, merging structures at lower redshifts continuously inject turbulence and drive a small-scale dynamo (see Sect.~\ref{sec:5}). This keeps the magnetic field strength relatively constant in both the ICM and ISM.

\end{enumerate}

The evolution in the non-radiative simulation (lower left panel) differs significantly. The magnetic field strength grows at a reduced rate at high redshifts. In phase \uproman{1}, the central regions are less dense (we show this in Fig.~\ref{fig:Profiles}) due to the lack of cooling, which reduces the efficiency of the dynamo. During phase \uproman{2} there are no central galaxies to develop and mix a strong magnetic field. Magnetic field growth only continues at an increased rate after the third phase, marked by major merger events, but still at a relatively low growth rate, compared to the radiative simulation. This results from the lower initial magnetic fields available for a dynamo, as there are no merging galaxies to provide their magnetic fields as an effective seed field. Notably, magnetic field growth accelerates only after the second significant merger event around $z \approx 3$, and saturates at $z \sim 1$.

\begin{figure*}
  \includegraphics[width=\textwidth]{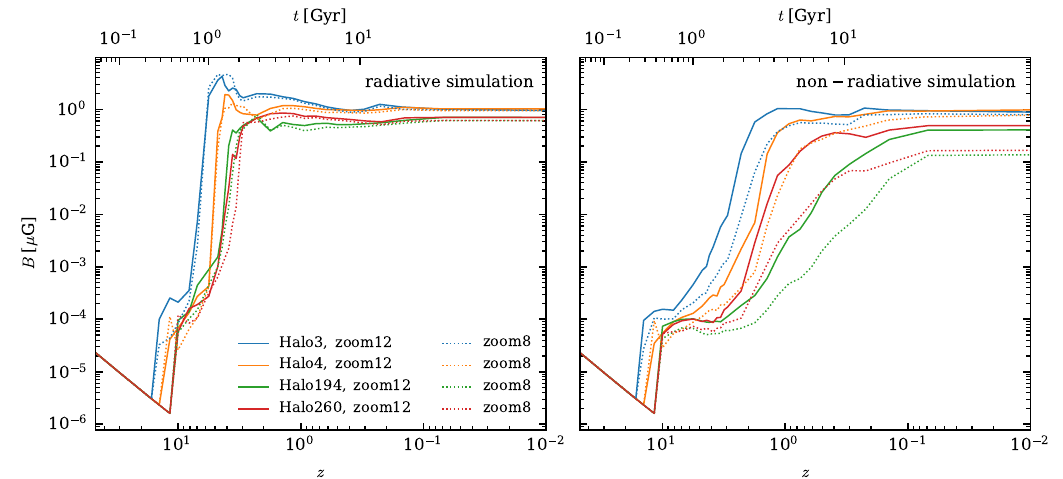}
  \caption{Resolution study of the evolution of the magnetic field strength averaged within $R_{200}$ for the radiative (left) and non-radiative (right) simulations. We present the evolution for all four halos (shown in different colors) and compare two resolution levels: ``zoom8'' and ``zoom12'' (see text). The radiative simulation shows no significant differences between the two zoom levels for all four halos, reflecting the fact that the magnetic field is amplified predominantly in the ISM, which is well resolved in both resolutions. In the non-radiative simulation, significant differences emerge between the two zoom levels, with zoom8 simulations showing smaller growth rates and lower final field strengths for the two smaller clusters.}
  \label{fig:B_time_resolution}
\end{figure*}

In the right panels of Fig.~\ref{fig:B_time}, we examine the magnetic field evolution in all four clusters. Magnetic field growth is more rapid in the radiative simulations, saturating at around $z \approx 4$ in the higher mass clusters, Halo3 and Halo4, and at around $z \approx 3$ in the lower mass clusters, Halo194 and Halo260. The saturation field strengths are $1~\upmu \mathrm{G}$ volume-averaged within $R_{200}$ in all four clusters. Notably, magnetic field evolution in the ICM of the radiative simulations closely mirrors that in the ISM, indicating a strong connection between the two. The ISM field saturates at around $20~\upmu \mathrm{G}$ in all halos. The ISM field in all halos saturates before the ICM. The ISM field in the most massive halo, Halo3 saturates slightly earlier in comparison to the other halos. In contrast, the non-radiative simulation achieves magnetic field saturation later, at around $z \approx 2$ in the higher mass halos and around $z \approx 1$ in the lower mass halos. The saturation field strengths are $1~\upmu \mathrm{G}$ in the higher mass clusters and a factor of two lower in the smaller clusters.

In Fig.~\ref{fig:B_time_resolution}, we address the important topic of numerical resolution on the evolution of the magnetic field strength. We compare two different resolution levels, `zoom8' ($8^3=512$ times better mass resolution compared to the parent simulation), and `zoom12' ($12^3=1728$ times better mass resolution compared to the parent simulation). As the numerical Reynolds number scales with the inverse cell size, this would give an eight times higher Reynolds number in the zoom8 simulation, compared to the parent simulation and a twelve times higher Reynolds number in the zoom12 simulation. The growth rate for the dynamo, in turn scales as $\Gamma \propto \mathrm{Re}^{1/2}$ for incompressible turbulence, which would result in a 2.8 (3.5) times higher growth rate in the zoom8 (zoom12) simulations, respectively. We find no difference between the magnetic field strength at different levels of resolution for adiabatic compression of the magnetic field due to protocluster collapse at $z\sim20$. In our radiative simulations, the saturated field is well converged while magnetic field growth is only somewhat slowed down in the lower resolution simulation during the dynamo growth phase. By contrast, in the non-radiative simulation the smaller numerical Reynolds number in combination with the finite time that a sizeable amount of turbulence is injected during merger events causes significantly slower dynamo growth. While the saturated magnetic field is just converged in the two massive clusters at $z=0$, it fails to converge in the two smaller clusters. In the following sections, we investigate the different growth phases in the radiative simulation in more detail, using Halo3 as our main example.

\subsection{Magnetic pre-enrichment in galaxies}
\label{sec:pre-enrichment}

As demonstrated, rapid magnetic field growth takes place in galaxies. Two stages are to be distinguished with regards to this magnetic pre-enrichment: We explore the influence of pre-enrichment in the BCG (phases \uproman{1} and \uproman{2}) and in merging substructures (phase \uproman{3} and \uproman{4}). This provides an interesting view of the magnetic field growth in galaxy clusters, showing that magnetic fields are intrinsically linked to its structure formation processes. 

\begin{figure*}
  \includegraphics[width=\textwidth]{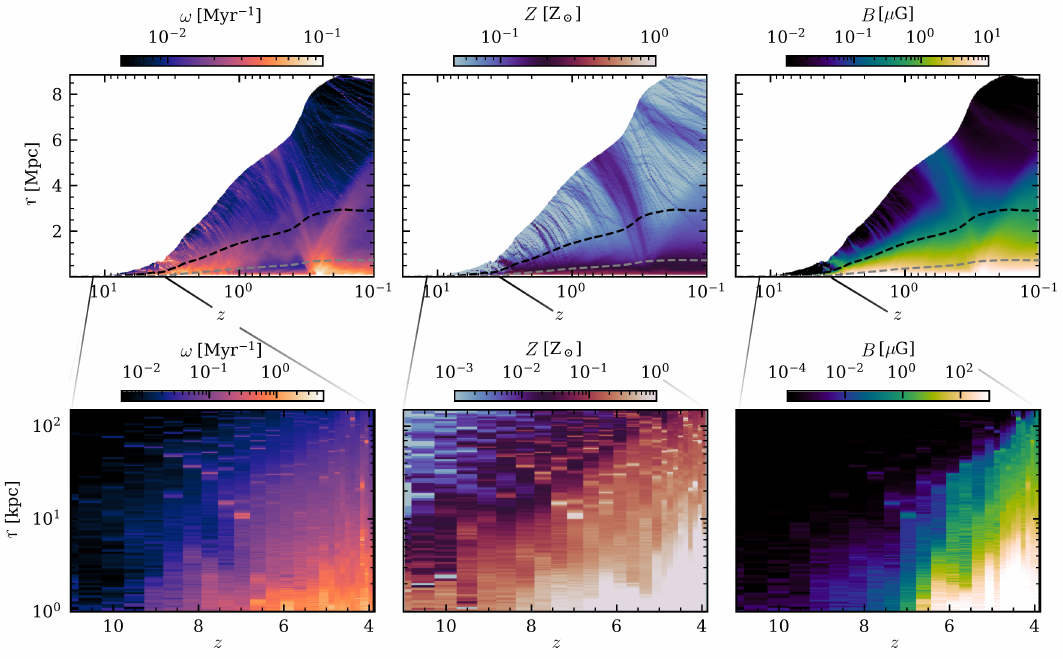}
  \caption{Redshift evolution of the physical radius corresponding to $3 R_{200}$ ($\approx9$~Mpc) surrounding the cluster in the radiative simulation of Halo3. From left to right, we color code the spherically averaged fields of vorticity, metallicity, and magnetic field strength. The upper row shows the redshift range $20>z>0.1$, highlighting the late time evolution dominated by mergers while the bottom row shows the redshift range $11>z>4$, focusing on the evolution in the central galaxy at higher redshifts. The dashed lines indicate the virial radius ($R_{200}$) and the central cluster region ($0.25  R_{200}$). We keep the maximum radius in the lower panels at a fixed physical value of $150~\mathrm{kpc}$ (see text). The lower panels show how vorticity and metallicity are enhanced in the center already at high redshifts, $z>9$. The central magnetic field amplification sets in shortly afterwards at $z \approx 7$. Turbulence from gravitational infall of cool gas amplifies the magnetic field and drives a small-scale dynamo. Stellar feedback additionally injects turbulence, distributes metals, and mixes magnetically enriched gas with less enriched gas. The upper panel highlights how merging substructures contribute their magnetically enriched gas to the host cluster. These merging structures also exhibit higher metallicity, indicating the connection between strong magnetic fields and the metal-rich ISM. }
  \label{fig:timeseries}
\end{figure*}

We show this in Fig.~\ref{fig:timeseries}, where we examine the time evolution of the spherically averaged radial profiles of the mass-weighted vorticity (left), the mass-weighted metallicities (middle) (indicative of stellar activity), and the volume-averaged root mean square magnetic field strength (right) of Halo3. In the upper panel, we focus on merger activity, showing the region $3 R_{200}(z)$ across the redshift range $20>z>0.1$. Dashed lines denote the virial radius ($R_{200}$) and the central cluster region ($0.25~R_{200}$), both in physical units. The lower panels focus on the redshift range $11>z>4$ to emphasize the early enrichment by metals and magnetic fields in the central galaxy and out to a fixed physical radius of $150~\mathrm{kpc}$, which corresponds to twice the virial radius of the central galaxy at $z=6.6$. 

First, we examine magnetic field growth in the central galaxy at higher redshifts in the lower panels. Initially, at $z>8$, magnetic field strength, metallicity, and vorticity have a low amplitude. At approximately $z \approx 9$, metallicity and vorticity notably increase to $Z = 1~\mathrm{Z}_\odot$\footnote{The metallicity is defined as the metal mass in a cell divided by the total cell mass $Z = M_Z/ M_\mathrm{total}$. We use $\mathrm{Z}_\odot = 0.0127$ for the solar metallicity, which means that $1.27 \%$ of the sun's total mass consists of metals.}, and $\omega = 10^{-1}~\mathrm{Myr}^{-1}$ in the central region, where the BCG is located. A strong magnetic field $B \approx 1~\upmu \mathrm{G}$ emerges slightly later at around $z \approx 7$ in the central region. This shows dynamo activity with a slight time lag following the onset of turbulence. Supersonic inflows of accreted gas generate turbulence in the cold, dense filaments, amplifying the magnetic field. Stellar feedback further interacts with the gas, distributing metals and creating additional turbulence, which mixes magnetized and less magnetized gas at larger radii. This leads to inside-out growth of the magnetic field, with initial amplification in the cluster center, followed by gradual enrichment of the outer regions.

The important role of merger activity is evident in the upper panels of Fig.~\ref{fig:timeseries}. Mergers appear as regions with increased magnetic field, metallicity, and vorticity beyond the virial radius, subsequently entering the cluster. Merging substructures exhibit high magnetic field and metallicity, emphasizing the role of stellar activity in redistributing the gas, and highlighting the correlation between metals and magnetic fields in intergalactic space. Mergers contribute their pre-enriched gas from within galaxies to the ICM of the main cluster.\footnote{Note that the contribution of individual galaxies to the magnetic pre-enrichment of the ICM changes over time: while at $z>8.5$ (phase \uproman{1}), nearly $90 \%$ of the gas mass inside $R_{200}$ is star forming, at $8.5>z>5.5$ (phase \uproman{2}), the mass of star forming gas decreases from $80 \%$ to $30 \%$ and drops below $4 \%$ until $z=3$ (phase \uproman{3}).} In Appendix~\ref{sec:density}, we present radial profiles showing how gas pre-processing in merging substructures enriches the cluster outskirts in both magnetic field strength and metallicity. 

Presumably, also ram pressure stripping plays an important role in removing the gas from infalling galaxies. This happens already at large radii, beyond the virial radius, as $\sim 90 \%$ of all galaxies arrive at the virial radius without any self bound gas at lower redshifts. Although the majority of the galaxies fulfills the \citet{1972ApJ...176....1G} criterion for ram pressure stripping to occur, this picture is probably too simplistic and can only give an upper limit because of insufficient numerical resolution to fully resolve this process \citep{2024MNRAS.527.5829S} and because of the pressurized effective equation of state, which responds differently to ram pressure in comparison to a multi-phase ISM. re-processing of metals and quenching of galaxies has been studied in observations and simulations of clusters and their vicinities, highlighting the role of ram pressure stripping and galactic winds  \citep{1984ApJ...285..426B, 2004PASJ...56...29F, 2012MNRAS.422.1609T, 2014A&A...570A..69B, 2014ApJ...781L..40E, 2015MNRAS.447..969B}. This process is observed not only near the cluster center but also at large radii, and prior to merging \citep{2004PASJ...56...29F,2013MNRAS.430.3017B, 2015ApJ...806..101H}. From Fig.~\ref{fig:timeseries} it is evident that the pre-processing also plays an important role for the magnetic field enrichment in galaxy clusters. 

\section{Magnetic field growth in the ISM and in the ICM}
\label{sec:4}

In this section, we distinguish between magnetic field growth in the ISM and ICM. At high redshifts, we show that a dynamo amplifies the magnetic field in the BCG and other early galaxies through compressible turbulence. Galactic winds transport and mix this magnetically pre-enriched gas into the ICM. In the ICM, a small-scale fluctuating dynamo driven by vortical turbulence further amplifies the magnetic field in the pre-enriched gas at low redshifts.

\subsection{Overview of fluctuating dynamos}
\label{ssec:overview dynamo}

First, we provide a high-level overview of the magnetic small-scale dynamo and the emergent scaling laws. A fluctuating small-scale dynamo operates as magnetic field lines stretch, twist, fold, and merge to amplify weak seed magnetic fields \citep{2005PhR...417....1B, 2018SSRv..214..122D, 2019JPlPh..85d2001R}. Such a dynamo operates in initially weakly magnetized, turbulent media like the ICM \citep{2017MNRAS.471.3212W, 2017MNRAS.464..210V, 2018SSRv..214..122D, 2020MNRAS.498.4983W} and exhibits exponential growth.

One sign of dynamo activity is the emergence of specific scaling laws that link magnetic power to wavenumber. In the kinematic phase of the dynamo, assuming an incompressible velocity field, \cite{1968JETP...26.1031K} demonstrated that magnetic power scales as $P(k) \propto k^{3/2}$ above the resistive scales. When the magnetic field becomes strong and back-reacts on the small scales, (see e.g. \citealt{1992ApJ...393..165V} and \citealt{2009ARA&A..47..291Z}), larger eddies shift the power spectrum peak to larger scales, following Kolmogorov scaling for incompressible turbulence \citep{1941DoSSR..30..301K}, $P(k) \propto k^{-5/3}$, on scales below the power spectrum peak. By doing so, also the largest magnetic correlation length of the dynamo amplified field shifts to larger scales. Theories and simulations show that the Kazantsev slope shifts to larger scales in this so-called non-linear regime \citep{PhysRevE.92.023010}. 

Studies have demonstrated that dynamo activity is also possible in compressible turbulence \citep{1985ZhETF..88..487K, 2005PhR...417....1B, 2012ApJ...754...99S, PhysRevE.92.023010}. Supersonically induced turbulence on large scales decays to incompressible turbulence below the sonic scale \citep{2020MNRAS.493.4400F, 2021NatAs...5..365F, 2023arXiv231203984B}. This can be understood such that compressible velocity fields caused by, for example, oblique shocks produce a non-negligible component of vortical turbulence (see e.g., \citealt{2018A&A...611A..15K, 2018AN....339..127K}, who investigated turbulence caused by clustered SN bubbles). This implies that the growth still takes place in Kolmogorov-type turbulence (with the addition of shock compression of magnetic fields), such that Kazantsev's theory is still applicable. According to simulations (see \citealt{2005PhR...417....1B} and references therein), a dynamo can develop, exhibiting Kazantsev scaling on large scales and Burgers scaling for compressible turbulence \citep{1969fecg.book.....B}, $P(k) \propto k^{-2}$, on smaller scales. However, the saturation strengths and growth rates are found to be lower, in comparison to Kolmogorov turbulence \citep{2005PhR...417....1B, 2012ApJ...754...99S, 2016JPlPh..82f5301F}. We will use this information to compare our results to the scaling laws described above in order to identify dynamo activity and in order to differentiate between dynamos induced by vortical or by compressive turbulence in Sect.~\ref{sec:power_spectra}.

The final saturation levels of the fluctuating small-scale dynamo are still not fully understood. Saturation is caused by the Lorentz force backreacting on the flow. This can be understood such that with increasing magnetic field strength, the backreaction of the Lorentz force onto the gas increases, causing an increasing velocity drift that lowers the effective magnetic Reynolds number \citep{1999PhRvL..83.2957S}. When it falls below the critical Reynolds number ($\mathrm{Re_c} \approx 60$) and the dynamo activity comes to a halt \citep{1999PhRvL..83.2957S}. It has been furthermore shown that the final saturation values of the magnetic-to-turbulent energy density depend on the magnetic Prandtl number and on the form of turbulence, where higher Prandtl numbers, as well as turbulence in vortical form yield higher saturation values, and lower Prandtl numbers and turbulence in compressive form yield lower saturation values \citep{2015PhRvE..92b3010S}.

There is ongoing debate about whether the dynamo always saturates at the same fraction of available kinetic energy (see, e.g., \citealt{2016IAUFM..29B.700M}). This idea stems from the fact that most kinetic and turbulent energy in the ICM arises from accretion and merger events. If the gravitational potential energy from a merger is converted into bulk and turbulent kinetic energy, the available kinetic energy should scale with the cluster mass. However, magnetic energy depends on density profiles and dynamo activity. Since density profiles in clusters are believed to be self-similar \citep{2017ApJ...843...28M}, magnetic fields may be comparable due to adiabatic gas compression. Dynamo activity, on the other hand, is driven by the amount of turbulent kinetic energy, meaning the ratio of magnetic to kinetic energy should reflect the dynamo's efficiency and be roughly universal across clusters \citep{2015Natur.523...59M}. We will investigate the efficiency of the dynamo amplification in our simulations in Sect.~\ref{sssec:saturation}.

\subsection{Power spectra for different regimes of magnetic field growth}
\label{sec:power_spectra}

\begin{figure*}
  \includegraphics[width=\textwidth]{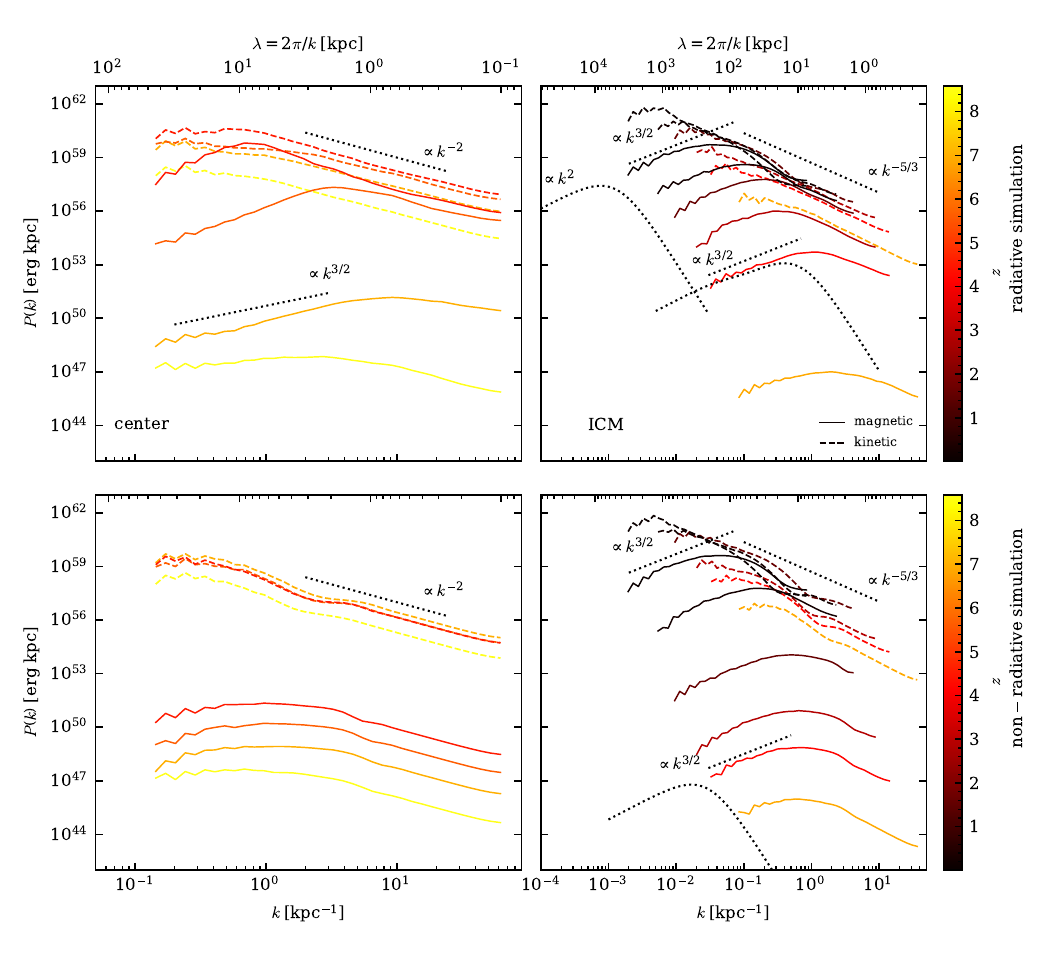}
  \caption{Kinetic (dashed) and magnetic (solid) power spectra of Halo3. The left panels show the power spectra in the \textit{central regions} within a fixed physical radius of $R_{200}(z=9)=20~\mathrm{kpc}$ for four different redshifts ($z=4.5, 5.5, 7, 8.5$) for both the radiative (upper) and the non-radiative (lower) simulations, aimed at tracing the high-redshift dynamo in the BCG. The right panels show the \textit{ICM} power spectra within $0.5 R_{200}(z)$, excluding the central $20~\mathrm{kpc}$, at six different redshifts ($z=0, 1.5, 2.5, 3.5, 4.5, 6.5$), focusing on dynamo activity in the ICM within the radiative and non-radiative simulations. Dotted lines illustrate different scaling behaviors (see text). The left panels show that a small-scale dynamo (with the Kazantsev scaling $\propto k^{3/2}$) driven by compressive Burgers turbulence ($\propto k^{-2}$) is only active in the central regions of the radiative simulation. By contrast, the right panels show the existence of a small-scale dynamo driven by incompressible turbulence ($\propto k^{-5/3}$) throughout the entire ICM within $0.5 R_{200}$ in both simulations, albeit at a faster rate in the radiative simulation. The contribution of a Lorentzian profile ($\propto k^2$ on scales larger than the scale height) at lower redshifts is minimal because the sharp cutoff at the characteristic exponential scale is located at rather large scales.}
  \label{fig:powerspec}
\end{figure*}

To differentiate between dynamos driven by compressible or vortical turbulence in the high/low redshift regime, we analyze power spectra. We have computed power spectra for Halo3 in order to connect its magnetic field evolution with the theory and literature outlined above. In Fig.~\ref{fig:powerspec}, we show the magnetic and kinetic power spectra across different redshifts ($8.5 > z > 0$) for Halo3. The power spectrum is computed from a zero-padded fast Fourier transform of the components of $\sqrt{\rho} \bm{\varv}$ and $\bm B/\sqrt{4 \pi}$. Calculations are performed in a sphere with a physical radius of $20~\mathrm{kpc}$ (which corresponds to the virial radius of the central galaxy at $z=9.5$) in the left panels, illustrating magnetic growth in the BCG, and a radius of $0.5  R_{200}$ in the right panels, excluding the central $20~\mathrm{kpc}$. We show the radiative simulation in the upper panels and the non-radiative simulation in the lower panels. Dotted lines represent various scaling behaviors: Burgers' model ($\propto k^{-2}$) for compressible turbulence, Kolmogorov's model ($\propto k^{-5/3}$) for incompressible turbulence, the Kazantsev model at large scales ($\propto k^{3/2}$) indicating dynamo activity, and a Lorentzian profile (that scales with $\propto k^2$ on large scales) as an indicator of an exponential magnetic field profile in real space.

First, we analyze magnetic power spectra in the central cluster regions at high redshifts ($8.5>z>4.5$) in Fig.~\ref{fig:powerspec} (left panels). Generally, magnetic energy in the radiative simulation exceeds that of the non-radiative simulation at all redshifts. At $z=8.5$, the magnetic energy remains low in both simulations. At $z=6.5$, the magnetic energy begins to grow on small scales in the radiative simulation, with the Kazantsev slope emerging on larger scales. The first peak of the magnetic power spectrum occurs at a characteristic scale of $2~\mathrm{kpc}$ in the radiative simulation and shifts to larger scales of $10~\mathrm{kpc}$ by $z=4.5$. At lower redshifts, the wavenumber scaling becomes somewhat steeper than the Kazantsev slope. On smaller scales, the spectrum follows the Burgers slope for compressible turbulence in both simulations. The compressible turbulence in the radiative simulation is due to gravitational, supersonic accretion of cold gas. In the non-radiative simulation the compressive turbulence is also due to cosmological accretion and mergers that move supersonically towards the center, but without cooling, which barely drives dynamo activity, as the eddy turnover times are larger. The turbulent energy is injected at larger scales ($\mathcal{L} \approx 10~\mathrm{kpc}$) as seen in the kinetic power spectra. 

The right panels of Fig.~\ref{fig:powerspec}, showing the entire ICM within $20 ~\mathrm{kpc} < r < 0.5 R_{200}$ over a broader redshift range $6.5>z>0$, demonstrate that magnetic energy increases over time, with its peak shifting to larger scales as expected in both simulations. In the non-radiative simulation, the magnetic power grows more continuously, whereas in the radiative simulation, it grows rapidly between $z = 8.5$ and $z=6.5$. While the power spectrum follows the Kolmogorov scaling $\propto k^{-5/3}$ on small scales it is ab initio unclear whether the cluster magnetic field profile or the Kazantsev scaling of the fluctuating dynamo sets the large-scale behavior of the power spectrum. 

We show in Appendix~\ref{sec:density} that the radial profiles of the magnetic field strength in the radiative and non-radiative simulations follow (double) exponential profiles, which implies (double) Lorentzian profiles in Fourier space. 
The associated one-dimensional power spectrum for a single-exponential profile is given by (Eq.~\ref{equ:lorentz2}):
\begin{align}
P(k) = 4 \pi  k^2 P_{3\mathrm{D}}(k)\propto \frac{\left( B_0 k_0 k \right)^2}{\left( k^2 + k_0^2 \right)^4},
\label{equ:lorentz2}
\end{align}
where $k_0^{-1}$ is the characteristic scale length of the exponential profile. This one-dimensional power spectrum scales $\propto k^2$ for small wave numbers and $\propto k^{-6}$ for large wave numbers. This implies that the contribution of the steeply rising Lorentzian form factor to the power spectra that we measure in our simulations on large scales could be flattened to the Kazantsev scaling $\propto k^{3/2}$ below the characteristic scale of the exponential profile because of the sharp drop-off of the squared Lorentzian towards small scales.  Most importantly, at lower redshifts, we expect the Kolmogorov spectrum to dominate for incompressible turbulence on scales smaller than the magnetic coherence length.

Figure~\ref{fig:powerspec} shows that over time, both the magnetic coherence scale and the characteristic scale length of the exponential magnetic profiles in galaxy clusters grow, with the exponential scale length growing faster than the magnetic coherence scale. While we show this for the radiative simulation, it is also true for the non-radiative simulation. On scales smaller than the exponential scale length (but still in the kinematic limit, i.e., on scales larger than the magnetic coherence scale), the logarithmic power spectrum slope is consistent with the Kazantsev spectrum, suggesting the emergence of an active small-scale dynamo in the ICM for redshifts $z\lesssim4.5$. By $z=0$, the exponential scale length of the magnetic profiles is two orders of magnitude larger than the magnetic coherence length, so that we can see a clear Kazantsev scaling on scales smaller than the magnetic coherence length in both simulations.

Notably, the injection scale of turbulence measured in the analyzed sphere is approximately the virial radius corresponding to the given time ($\mathcal{L} \approx R_{200}$). Most of the turbulent energy comes from mergers that excite supersonic velocities beyond the virial radius and transitioning to subsonic turbulence when entering the virial radius, minor contributions are stemming from stellar and AGN feedback. 

\begin{figure*}
  \centering\includegraphics[width=0.9\textwidth]{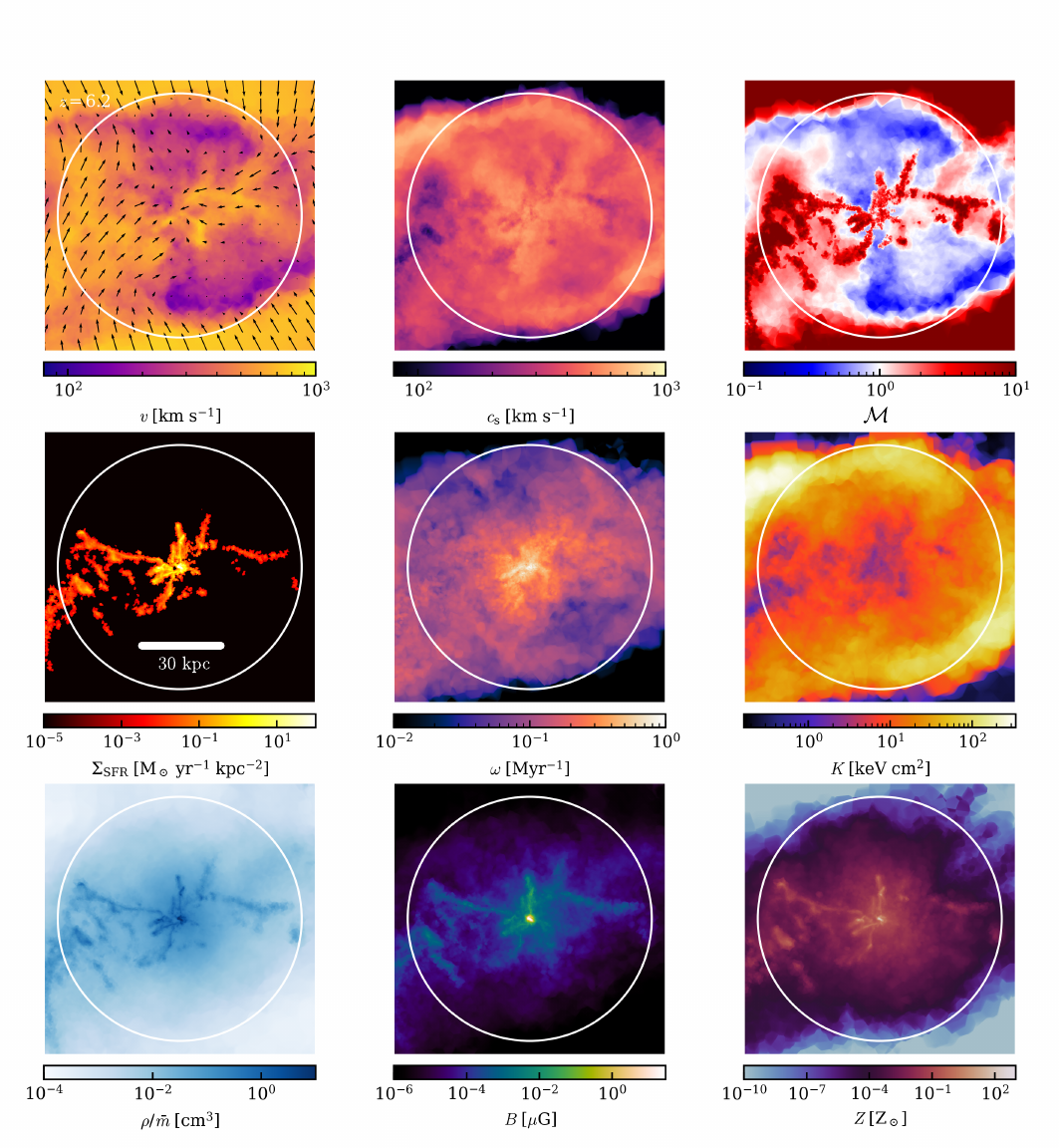}
  \caption{Thin projections through the center of Halo3 in the radiative simulation with sidelength $2.2 R_{200}$ at $z=6.2$ and thickness $0.5R_{200}$. From top left to bottom right, the nine panels display gas velocity in the cluster rest frame, sound speed, Mach number, SFR surface density, vorticity, cluster entropy, particle number density, magnetic field strength, and metallicity. The velocity field is shown together with its absolute value in the first panel. Regions with cold, supersonic gaseous inflows create vorticity and display strong magnetic fields. Star formation in the high density regions creates stellar winds that interact with the inflowing gas and produce vorticity. Given the shallower entropy profile at larger radii, this vorticity efficiently mixes highly magnetized gas with weaker magnetized gas, distributing metals across the cluster.}
  \label{fig:feedback_projection}
\end{figure*}

\subsection{Magnetic field growth in the ISM}
\label{sec:ISM dynamo}

At high redshifts, $z>5.5$, most of the ICM gas is star-forming, with magnetic field growth primarily occurring in the ISM. In this section, we demonstrate that the turbulence required for the dynamo process in the ISM is driven by supersonic cooling flows, while stellar and AGN feedback generate magnetized outflows that effectively enrich the remaining cluster volume with magnetic fields.

\subsubsection{Gravitationally driven dynamo, and stellar and AGN feedback-induced mixing}
\label{ssec:winds}

\begin{figure*}
  \includegraphics[width=\textwidth]{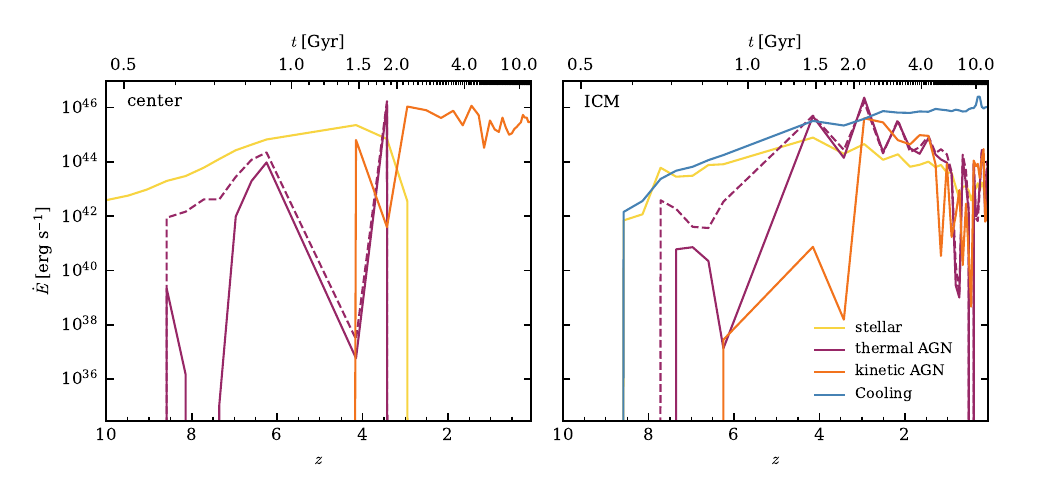}
  \caption{Energy injection rates into gas cells of Halo3 in the radiative simulation, differentiating between feedback that is launched from the BCG center ($r<20~\mathrm{kpc}$)(left panel), and feedback that is launched from the entire cluster excluding the center (right panel). We present different energy injection rates that originate from stellar feedback (yellow), as well as thermal (purple) and kinetic (orange) AGN feedback. The injected thermal AGN feedback (dashed purple) is numerically reduced in star-forming cells where an effective equation of state is enforced, implying a rapid energy loss and a smaller effective thermal AGN feedback energy rate (solid purple). Radiative cooling losses (blue) are displayed for the ICM in the right panel. At high redshifts, both in the BCG and the entire cluster, the primary energy injection rate is due to stellar feedback. Subsequently, AGN feedback also becomes a significant contributor.}
  \label{fig:feedback}
\end{figure*}

Figure~\ref{fig:feedback_projection} presents thin projections from the radiative simulation of Halo3, showing gas speed, thermal sound speed, Mach numbers, star formation rate (SFR) surface density, vorticity, entropy $K=k_\rmn{B} T n^{-2/3}$, particle number density, magnetic field strength, and metallicity at $z=6.2$. This analyzed time lies in phase \uproman{2}, during which the magnetic field is mainly amplified in the BCG and then partially mixed with the surrounding ICM. From the velocity panel it is evident that there is a gravitationally driven gas inflow towards the center. The gas flows along dense, star-forming filaments with a high velocity, which creates strong shocks in these cold filaments. Shocks encountering density inhomogeneities produce a high amount of vorticity. Due to the high densities, the resolved eddy turnover rates are very high and by such, enable efficient dynamo amplification. 

Stellar winds are launched from regions with a high SFR, which drives galactic winds that impact the inner cluster region ($r\lesssim0.5 R_{200}$). These produce turbulence and can interact with the infalling gas, as can be seen in the upper left corner of the velocity panel, close to the virial radius. Additionally, the gravitational potential of the forming cluster accretes gas that forms a shock close to the virial radius upon encountering the dense ICM. The inhomogeneous density distribution causes curved virial shocks, which generates vortical turbulence according to Crocco's theorem. The turbulence stemming from both sources can efficiently mix stronger magnetized gas from lower radii with less magnetized gas at higher radii due to the shallow entropy profile in the cluster outskirts (see the middle right panel and the shallow entropy profiles at larger radii in Appendix~\ref{fig:Profiles}). Metals are widely distributed throughout the entire ICM which is a testament to the effective turbulent mixing due to stellar winds, as shown in the lower right panel. 

We investigate the main energy injection sources into the gas to identify drivers of galactic winds that mix and distribute high magnetic field gas. In Fig.~\ref{fig:feedback}, we show the energy injection rates of different feedback channels in the radiative simulation of Halo3: stellar feedback, thermal AGN feedback, kinetic AGN feedback, and compare this to the energy loss rate due to cooling. In the left panel of Fig.~\ref{fig:feedback} (representative of the BCG), we display the energy injection rates that are launched from the central $20~\mathrm{kpc}$, while the right panel shows the energy injection rates that are launched from the remaining part of the cluster (excluding the center) within the virial radius. Our calculations follow the methodology outlined in \cite{2018MNRAS.479.4056W} (see their section~2.3 for AGN feedback and their section~2.4 for stellar feedback, the models used to simulate AGN and stellar feedback are described in Sect.~\ref{sec:illustris-tng}). 

As observed in the left panel, stellar feedback serves as the primary energy injection source launched from the center until approximately $z \approx 6$. Subsequently, there is a brief phase ($5>z>3$) where also thermal AGN feedback becomes a main contributor while kinetic AGN feedback starts to dominate the energy injection after $z=3$. Examining the right panel of Fig.~\ref{fig:feedback}, we discern two distinct phases of feedback injection in the entire ICM: at $z>5$, stellar feedback is the primary energy source, while at $z<5$, contributions from AGN, both thermal and kinetic, dominate the energy injection. The feedback acts in such a way that it removes gas from galaxies and mixes it into the ICM. Apart from the feedback channels listed above, cluster galaxies, being located in high-density regions, experience gas loss due to ram pressure stripping during and after phase \uproman{3} (see Sect.~\ref{sec:pre-enrichment}).

\subsubsection{Adiabatically expanding galactic magnetic fields into the ICM}

\begin{figure*}
  \includegraphics[width=\textwidth]{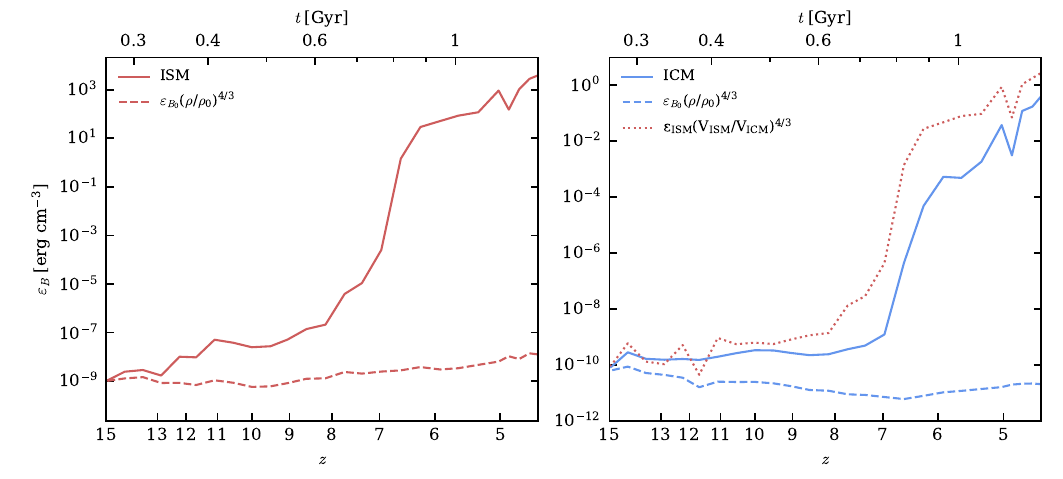} 
  \caption{Left panel: The magnetic energy density in the ISM (red). We measure the energy density in an expanding sphere with a radius equal to the virial radius at the time. When the magnetic energy density inside this sphere has, for the first time, grown by a factor of $10^2$ compared to the adiabatically compressed field right after the protocluster collapse, we fix the radius at this redshift. Right panel: The magnetic energy density in the ICM (blue) is compared to the magnetic energy density expected from expanding the ISM magnetic field purely adiabatically to the volume of the ICM (red dotted). This estimate even overshoots the total magnetic energy in the ICM. The lower measured ICM magnetic field strengths, compared to adiabatically expanded ISM fields, may result from depletion during star formation, retention in the ISM, or unaccounted mixing losses.}
  \label{fig:a}
\end{figure*}

We demonstrate that as galaxies fall into the cluster potential, magnetized outflows and ram pressure stripping significantly contribute to the magnetization of clusters. We do this by demonstrating that the simulated magnetic field strength in the ICM can be largely explained by adiabatically expanding the magnetic field from galaxies. We first calculate the volume of the ISM, $V_\mathrm{ISM}$, and the mean magnetic energy inside of it, $\varepsilon_\mathrm{ISM}$. We then imagine that the ISM volume is adiabatically and isotropically expanded to fill the ICM volume while conserving magnetic flux. Because $B\propto \rho^{2/3}$ during this process, the magnetic energy scales as $\varepsilon_B\propto B^2 \propto \rho^{4/3} \propto V^{-4/3}$ and we have $\varepsilon_B V^{4/3} = \mathrm{const}$. We can use this relation to make a prediction for the mean magnetic energy density in the ICM if it has been purely generated as a result of adiabatic expansion of the ISM:
\begin{align}
    \varepsilon_{\textrm{ICM,\,expanding\,ISM}} = \varepsilon_\mathrm{ISM} \left(\frac{V_\mathrm{ISM}}{V_\mathrm{ICM}}\right)^{4/3}.
     \label{equ:h1}
\end{align}
We can also calculate the magnetic energy as a result of compressing/expanding the pre-collapsed protocluster magnetic field and density to its current density, which yields
\begin{align}
    \varepsilon_{\textrm{ICM,\,collapsing\,cluster}} = \varepsilon_{B,0} \left(\frac{\rho}{\rho_0}\right)^{4/3},
    \label{equ:h2}
\end{align}
where $\varepsilon_{B,0}$ and $\rho_0$ are the mean magnetic energy density and density at $z=15$, respectively.

Figure~\ref{fig:a} illustrates the evolution of magnetic field energy ty between $z=15$ and $3.5$ 4.5 Halo3. Solid lines represent the measured mean energy density with a volume (as explained below) while dashed lines show the expectations as a result of adiabatically changes of the gas from the initial magnetic field strength and gas density after the protocluster collapse (equation \ref{equ:h2}). The radius of the measurement volume expands as the halo's virial radius. Once the energy density within this radius exceeds the expectations from adiabatic compression for the first time by a factor of $10^2$, we set the virial radius at this time as a fixed radius for further measurements, following the method of \cite{2024MNRAS.528.2308P}. This radius is $R_{200} = 74~\mathrm{kpc}$ at $z \approx 6.5$. The left panel displays magnetic energy densities for the ISM (red), while the right panel shows the magnetic energy density for the ICM (blue). Additionally, we represent the adiabatically expanded magnetic field from the ISM in the right panel with dotted red lines (equation \ref{equ:h1}).

From the left panel of Fig.~\ref{fig:a}, it is evident that a dynamo is operational, causing the magnetic field to surpass the expectations from adiabatic compression by multiple orders of magnitude and amplifying the magnetic field exponentially. In the right panel of Fig.~\ref{fig:a}, the energy density in the ICM similarly exceeds that achievable through adiabatic compression alone by several orders of magnitude.However, expanding the ISM magnetic field into the ICM overshoots the actual values in the ICM.\footnote{Note that by $z=4.5$ still $30 \%$ of the total gas mass in the cluster is star forming.}  This discrepancy can arise because this estimate does neither account for depletion of the magnetic field during star formation events nor for the magnetic field that is retained in the ISM. Note that this argument assumes full mixing of the adiabatically expanded magnetic field with the weaker magnetized ICM as well as no magnetic dissipation during this process, which will also lead to estimates being higher than the true value. 

Clearly, all these effects can decrease the magnetic field strengths in the ICM below the measured values, thereby requiring a small-scale dynamo to amplify this already pre-enriched magnetic field (see Sect.~\ref{sec:5}). In any case, the pre-enrichment scenario obviates the need for high dynamo growth rates in the ICM to explain a strong magnetic field at high redshift. Interestingly, the adiabatic evolution of the initial magnetic field shows little evolution as adiabatic dilution due to cosmic expansion is approximately balanced by adiabatic compression due to gravitational collapse.

\subsubsection{Growth rates in the ISM}
\label{sssec:growth}

\begin{figure*}
  \includegraphics[width=\textwidth]{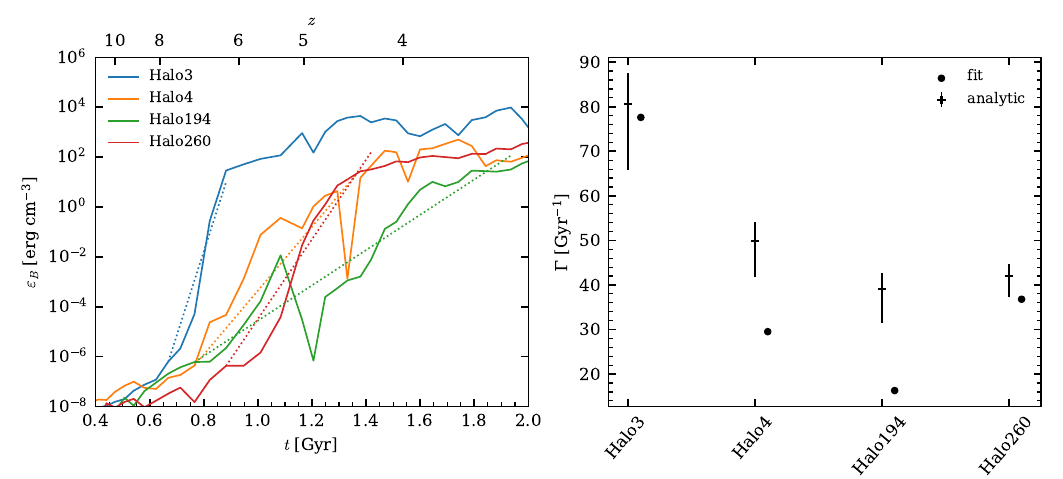}
  \caption{Left: Time evolution of the magnetic energy density with fits of the exponential growth rates for the ISM in all four halos (dotted lines). The fits start when magnetic energy density has for the first time grown to a value at or above $4\times10^{-7}~\mathrm{erg~cm}^{-3}$ and ends when the magnetic energy density has for the first time reached a value of at least $40~\mathrm{erg~cm}^{-3}$. We subtract the contributions from adiabatic compression, following the relation given in equation~\eqref{equ:h2}, where we fix $B_0$ and $\rho_0$ to be the measured values after the cluster collapse, and $\varepsilon_B(t)$, $\rho(t)$ are measured in each snapshot. In the right panel, we display the fitted growth rates (circles) along with the expected growth rates (crosses) that include error bars as described in the text. Our analytic and measured growth rates agree within a factor of two, with the expected growth rates being slightly larger than the measurements.}
 \label{fig:b}
\end{figure*}

Figure~\ref{fig:b} shows the evolution of the magnetic energy density in the ISM of all four halos (left panel). The contributions due to adiabatic compression of the gas have been linearly subtracted in order to isolate the amplification via dynamo activity. A fluctuating small-scale dynamo shows a characteristic exponential growth:
\begin{equation}
\varepsilon_B(t) = \varepsilon_{B_0} \mathrm{e}^{\Gamma t},
\label{equ:B_exp}
\end{equation}
where $\varepsilon_{B_0}$ is the initial magnetic field strength and $\Gamma$ is the growth rate. We fit magnetic growth rates in the ISM of all four halos (shown as dotted lines) and compare these fitted growth rates against analytical growth rates. The fit to the curves is a least square fit minimizing $(\Delta \log \varepsilon_B)^2$ and it is performed linearly in time. We restrict the fit to the interval between the time when the renormalized magnetic energy density first reaches a value of $4 \times 10^{-7}~\mathrm{erg~cm}^{-3}$ and the time when it first exceeds a value of $40~\mathrm{erg~cm}^{-3}$. The fits for Halo3 and Halo260 approximate the overall growth of magnetic energy density in the ISM very well. In contrast, the fits for Halo4 and Halo194 are less accurate due to more spiky and irregular growth patterns, likely resulting from feedback events that eject magnetized gas from the ISM or lock it up in stars. Additionally, the ISM likely contains multiple superimposed dynamos \citep{2022MNRAS.515.4229P}, making it difficult to fit a single growth rate. 

In order to compare our measured growth rates against theoretical models, we adopt the equation of \citet{2012PhRvE..85b6303S}:\footnote{This equation was derived in the $\mathrm{P_m} \rightarrow \infty$ limit, where $\mathrm{P_m}$ is the magnetic Prandtl number. Calculations for $\mathrm{P_m} \ll 1$ show that the same equation applies when $\mathrm{Re}$ is exchanged with $\mathrm{Re_m}$ \citep{2012PhRvE..86f6412S}. For the $\mathrm{P_m} \approx 1$ regime, that holds true in our simulations, there is no explicit expression. We therefore adopt the equation given by \citet{2012PhRvE..85b6303S} and show that the results are comparable to measured growth rates in our simulations.}
\begin{align}
     	\Gamma=\frac{163-304 \theta}{60}\frac{\mathcal{V}}{\mathcal{L}}\mathrm{Re}^{(1-\theta)/(1+\theta)},
     	\label{equ:growth} 
\end{align}
where $\mathrm{Re}$ is the Reynolds number, $\mathcal{V}$ is the injection velocity of the turbulence, and $\mathcal{L}$ is the injection scale of the turbulence. The parameter $\theta$ is characterized by the type of turbulence, with $\theta=1/3$ for Kolmogorov and $\theta=1/2$ for Burgers turbulence. Because we encounter compressible turbulence in the ISM, we use $\theta=1/2$. 

We use an injection scale of $\mathcal{L} = 10~\mathrm{kpc}$, as observed in the kinetic power spectra. The injection velocity $\mathcal{V}$  is calculated as the mass-weighted root mean square velocity of all cells in the center of mass frame. This approach, is based on the assumption that turbulence originates from various sources such as star-forming regions, AGN feedback, and turbulence injected in the outskirts that cascades towards the center. We assume that these different sources of turbulence are well mixed in the center and that the mean velocity is somewhat connected to the injection velocity. The numerical Reynolds number can be estimated via \citep{2022MNRAS.515.4229P}:
\begin{align}
\mathrm{Re}_\mathrm{num} = \frac{\mathcal{L} \mathcal{V}}{\nu_\mathrm{num}} \sim \frac{3 \mathcal{L} \mathcal{V}}{d_\mathrm{cell}\varv_{\mathrm{th}}}.
\label{equ:reynold}
\end{align}
Here, $\nu_\mathrm{num}$ is the numerical viscosity, and $d_\mathrm{cell}$ is the diameter of the cell which we calculate from the cell volume, assuming the cell is spherical. The velocity $\varv_\mathrm{th}=\sqrt{2 k_\mathrm{B}T /[\bar{m} (\gamma -1)] }$ is the thermal velocity of the cell. For $\varv_\mathrm{th}$, we adopt the mass weighted thermal velocity of all star forming cells that lie within the measured radius. We take the median cell diameter and set the 25th percentile and 75th percentile cell diameters as upper and lower uncertainty regions.

In the right panel of Fig.~\ref{fig:b}, we compare the analytical growth rates making use of equations~\eqref{equ:growth} and \eqref{equ:reynold} with the fitted growth rates obtained from the left panel. We find that the fitted and calculated growth rates closely align for all halos. Generally, the fitted growth rates are lower than the analytical ones, possibly due to losses as a result of the locking up of the magnetic field during star formation. For Halo3 and Halo260, where our fits approximated the growth very well, the fitted and analytic growth rates are close to each other. For Halo4 and Halo194, where our fits did not perform as well, the fitted growth rates are lower (by factors of around two) than the actual measured ones.

\begin{figure*}
  \includegraphics[width=\textwidth]{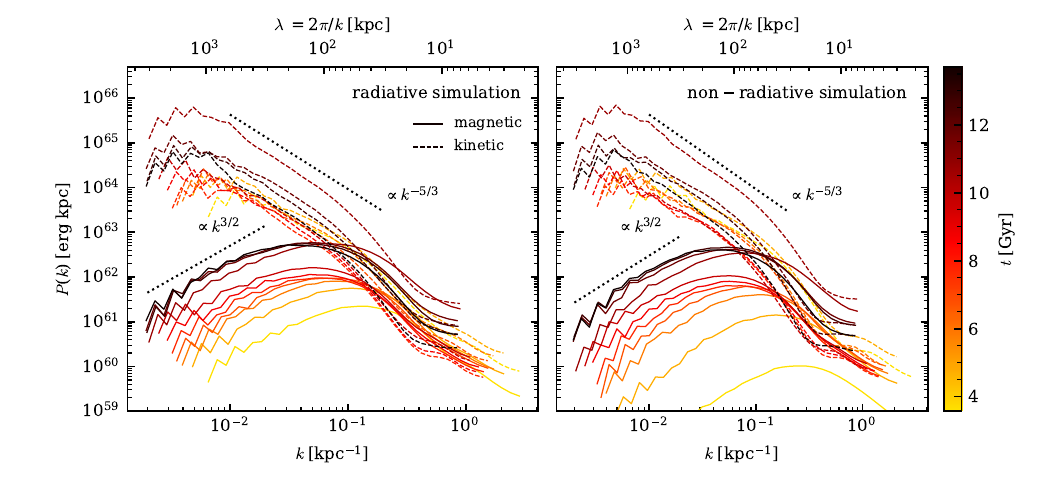}
  \caption{Kinetic (dashed) and magnetic (solid) power spectra of Halo3 are presented within a radius of $0.5 R_{200}$ at $11$ different cosmic times $t$ that are equally spaced ($3.6~\mathrm{Gyr} < t < 13.7~\mathrm{Gyr}$ with approximately $1~\mathrm{Gyr}$ separation in between; this corresponds to $1.5>z>0$) for both the radiative (left) and the non-radiative (right) simulations. These times correspond to our phase \uproman{4} (defined in Fig.~\ref{fig:B_time}), where the small-scale dynamo operates in the ICM. This fluctuating dynamo amplifies the field ejected from galaxies and grows it in strength and correlation scale via turbulence injected by cluster merger events. This process also occurs in the non-radiative simulations, albeit at a much reduced level because in this model there is no galactic pre-processing of magnetic fields. Over time, the peak magnetic energy shifts to larger scales.}
  \label{fig:powerspec2}
\end{figure*}

\subsection{Magnetic field growth in the ICM}
\label{sec:5}At lower redshifts ($z<3.5$), the cluster has significantly grown, reducing the impact of individual galaxies on magnetic pre-enrichment. In this section, we provide further evidence for dynamo activity within the ICM.

\subsubsection{Dynamo amplified magnetic correlation lengths}

The magnetic field strength in the ICM remains constant during phase \uproman{4}, as seen in Fig.~\ref{fig:B_time}. However, the accretion of less magnetized gas, as well as the decay of the magnetic field, would lower the field strength during this phase. The fact that it stays constant instead, shows that there are additional dynamo processes that continuously grow the field. The power spectra in Fig.~\ref{fig:powerspec2} are  calculated within a sphere of $0.5 R_{200}$ and illustrate the growth of magnetic energy over the last $10~\mathrm{Gyr}$ in both radiative and non-radiative simulations. As predicted by dynamo theory, the peak of magnetic energy shifts to larger scales. The similarity in the evolution of the magnetic correlation length in both simulations indicates that kinetic energy from cosmic accretion and merger events that are present in both cases, is the source. A major merger event at $t=10.7~\mathrm{Gyr}~(z = 0.3)$ moves the peak magnetic energy to smaller scales, where smaller-scale turbulence effectively amplifies the fields. Subsequently, the peak shifts back to larger scales while the total magnetic energy remains nearly constant. At $z=0$, the largest dynamo amplified correlation length inside $0.5 R_{200}$ is $\approx$$200~\mathrm{kpc}$. The kinetic power spectra follow a Kolmogorov scaling on larger scales ($10^{-2} \,\mathrm{kpc}^{-1} \, < k < 10^{-1} \, \mathrm{kpc}^{-1}$) but steepen at smaller scales. At even smaller scales, we see a flattening that could indicate an additional injection of small-scale kinetic energy, e.g., as a result of feedback; a phenomenon that can be studied in a future investigation. The magnetic power continues to grow together with the cluster, as (less magnetized) mergers bring new mass and kinetic energy, which acts to keep the magnetic field strength constant in the growing cluster volume.

\subsubsection{Phase space diagrams}
\label{ssec:PSDig}

\begin{figure*}
  \includegraphics[width=\textwidth]{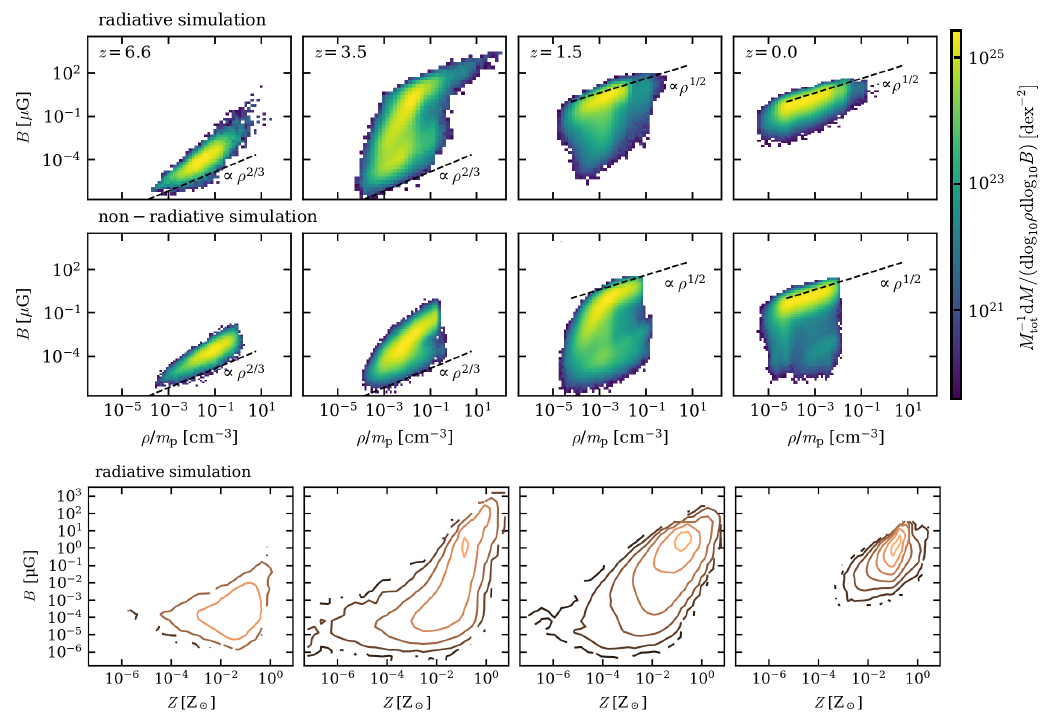}
  \caption{Mass weighted histograms in the magnetic field vs.\ gas density plane (upper two rows) in the ICM. Adiabatic compression of a flux-frozen magnetic field implies $B \propto \rho^{2/3}$, and a self-gravitating system with a saturated dynamo scales as $B \propto \rho^{1/2}$ (see black dashed lines). The radiative simulation starts to deviate early and first at high densities (corresponding to small scales) from the adiabatic relation, hinting towards dynamo activity at high redshifts. A rapid growth and early saturation is visible. By contrast, in the non-radiative simulation, magnetic amplification is significantly delayed and saturation is only reached at $z=0$. The lower row shows the magnetic field vs.\ metallicity histogram in the ICM. The contours of constant density are indicated by brown-orange lines. These panels illustrate how the magnetic field is first amplified in high metallicity gas (in the center) and then, subsequently, in lower metallicity gas. The correlation between both quantities highlights the contributions of pre-enriched gas (in magnetic fields and metals) from merging galaxies in enriching the outskirts with both magnetic fields and metals.}
  \label{fig:Phase}
\end{figure*}

To further investigate the magnetic field growth through the kinematic and saturated phases of the dynamo, we analyze the phase space diagrams of Halo3 in the upper two rows of Fig.~\ref{fig:Phase}. These diagrams show mass-weighted histograms in the two-dimensional plane spanned by magnetic field strength and density for four distinct redshifts. 

Adiabatic compression in an isotropically collapsing halo and magnetic flux freezing yields the scaling $B \propto \rho^{2/3}$. Observations of nearby clusters revealed a scaling of $B \propto \rho^{0.47}$ \citep{2010A&A...513A..30B,2017A&A...603A.122G}. Indeed, in a saturated small-scale dynamo, the specific magnetic energy $\mathcal{E}_B\propto \varv_\rmn{A}^2=B^2/(4\pi\rho)$ remains constant provided the turbulent energy does not change, which implies $B \propto \rho^{1/2}$  \citep{2020ApJ...899..115X}. Thus, we adopt this scaling for the saturated dynamo stage.

At $z=6.6$, the high-density region (with the shortest dynamical time) in the radiative simulation first departs from the adiabatic relationship, hinting at dynamo activity or mixing of the ICM with magnetically enriched ISM. By $z=3.5$, the cluster experiences major merger series, which causes the entire distribution in the radiative panel to deviate from the adiabatic expectation, especially at higher densities. In the non-radiative simulation, the denser region begins to gradually diverge from adiabatic behavior. At $z=1.5$, the distribution in the radiative simulation enters the saturated regime, while the non-radiative simulation exhibits two power-law segments. These may represent the influence of two superposed dynamos, possibly triggered by the two major merger events. By $z=0$, both distributions closely adhere to the scaling $B \propto \rho^{1/2}$ characteristic of a saturated dynamo. In the non-radiative simulation, there are some regions that are not yet saturated, even though most of the ICM is in the saturated regime. These observations summarize our findings: 1) magnetic fields grow initially via adiabatic compression, 2) there is a small-scale fluctuating dynamo that starts acting in the densest regions with the fastest eddy turnover rates and which is clearly much faster in the radiative simulation, and 3) eventually both simulations enter the saturated dynamo regime.

In the lower panel of Fig.~\ref{fig:Phase}, 2D histograms of magnetic field strength and metallicity are presented for the entire ICM (excluding star forming regions). Density contours are indicated with brown-orange lines. In our model, metals are transported from the ISM to the ICM in form of stellar feedback driven outflows. While in reality the outflows also carry magnetic fields, this does not happen in our subgrid model, therefore our results represent lower limits to the ICM magnetization at high redshifts. At $z=6.6$, adiabatic compression of the proto-galaxy has increased the magnetic field to $\sim10^{-4}~\upmu$G while the magnetic field in the ICM starts to be amplified by a dynamo in the central regions that have been polluted with metals. By $z=3.5$, the cluster has experienced many merger events with galaxies that contribute their high metallicity, strong magnetic field gas to the ICM yielding a positive correlation between both quantities. Additionally, the subsonic dynamo in the ICM starts to operate and magnetically amplifies the lower-metallicity gas, resulting in saturated magnetic fields by $z=0$.

\subsubsection{Saturation strengths}
\label{sssec:saturation}

\begin{figure}
  \includegraphics[width=\columnwidth]{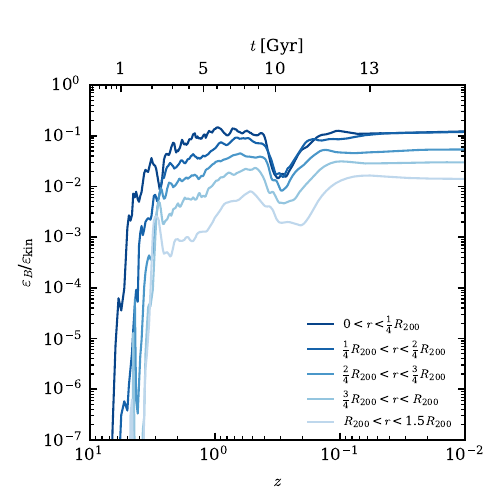}
  \caption{Ratios of the magnetic-to-kinetic energy density in the ICM calculated by averaging within spherical shells for the radiative simulation of Halo3. The progressively lighter colors indicate radial shells that move to larger radii. The magnetic energy saturates at around $10\%$ in the innermost bin.  The constant ratios at low redshifts indicate dynamo activity, while the lower values in the outer radial bins are likely due to a higher proportion of bulk kinetic in comparison to turbulent energy, as well as weaker magnetic fields in lower-density gas (see text).}
  \label{fig:Energies}
\end{figure}

Another piece of evidence for a small-scale dynamo in the saturated regime is the constant ratio between magnetic and kinetic energy. There is an ongoing debate surrounding the existence of a universal saturation ratio between magnetic field energy density and kinetic energy density, both of which are expected to scale with cluster mass (see Sect.~\ref{ssec:overview dynamo}).

In Fig.~\ref{fig:Energies}, we present magnetic-to-kinetic energy density ratios for various radial bins (indicated by blue shadings) of Halo3. The kinetic energy is the sum of kinetic energies of each cell in a radial bin in the center of mass system. The ratio stays constant with time for all radial bins, indicating the presence of a dynamo. The ICM in the innermost radial bin ($0<r<0.25R_{200}$) saturates at $10\%$ of kinetic energy, and the outermost radial bin ($R_{200}<r<1.5R_{200}$) at $1\%$. The reason for the decreasing magnetic-to-kinetic energy density ratio towards the outer radial bins can be twofold. 1) There is a progressively larger fraction of bulk kinetic energy in comparison to the turbulent energy density at larger radii. 2) The saturated field strength of a small-scale dynamo scales as $B\propto\rho^{1/2}$ (see Fig.~\ref{fig:Phase}), implying lower magnetic field strengths at larger radii (see Fig.~\ref{fig:Profiles}). A merger causes the magnetic-to-kinetic energy ratio to drop at all radii at $z \sim 0.3$ due to an injection of kinetic energy and gas with a lower magnetization into the host halo. We intend to improve the turbulent kinetic energy identification in future work.

\subsection{A magnetic dynamo on magneto-hydrodynamics scales}

\begin{figure}
  \includegraphics[width=\columnwidth]{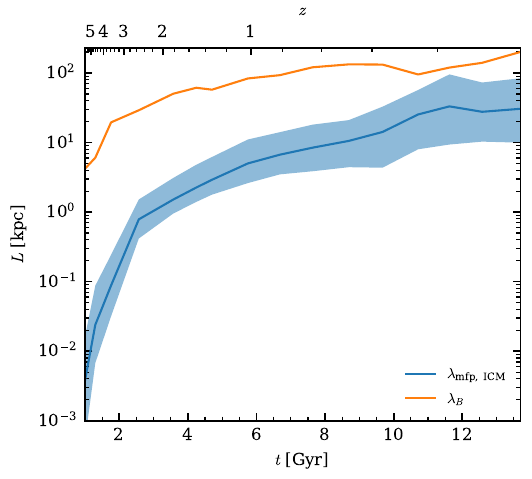}
  \caption{A comparison of the particle mean free path in the fully ionized ICM (equation~\ref{equ:mfp}) and the magnetic coherence length (i.e., the peak of the magnetic power spectrum from Fig.~\ref{fig:powerspec}). This shows that by the time exponential amplification in the ICM begins (at $z=5.5$), the magnetic coherence length has already significantly increased as a result of the dynamo activity in the ISM and the successive transport of the magnetic field into the ICM by means of galactic winds or ram pressure stripping, which further expanded the coherence length. This demonstrates that the dynamo in the ICM operates in the fully collisional regime throughout this process, rendering our MHD approximation valid.}
  \label{fig:MFP}
\end{figure}

The question arises whether the use of the MHD approximation is justified, as it assumes that the scales on which the small-scale dynamo operates in our simulation are collisional. This condition can be refined by requiring that the magnetic coherence length peak exceeds the particle mean free path during magnetic field amplification. Here, we demonstrate that in our proposed two-stage model of magnetic field evolution -- high redshift growth in the ISM and low redshift growth in the ICM -- amplification occurs entirely on collisional scales.

To start, we need to assume a generation process of magnetic fields. There are several processes that produce macroscopic fields on kpc scales or larger, including cosmological phase transitions in the early universe \citep{2002RvMP...74..775W,2013A&ARv..21...62D} or the Biermann battery mechanism in ionization fronts or at shocks of strong, supernova-driven outflows \citep{2021MNRAS.504.2346A}. Turbulent motions then redistribute and amplify this seed field via dynamo activity.

In our proposed picture, field growth advanced through the following three stages:
\begin{enumerate}
\item In Section~\ref{sec:ISM dynamo}, we demonstrated that at high redshifts (between $z=8.5$ and $5.5$), the magnetic field is primarily amplified in the ISM, which is well approximated by the MHD approximation (see e.g. \citealt{Schekochihin2009ApJS}). That is, the mean free path for typical ISM values is 1 AU in the cold neutral medium \citep{Ryden2021}  and $\approx 0.05$ AU in the warm ionized medium \citep{Schekochihin2009ApJS}, which are much shorter than our numerical resolution. The dynamo exponentially amplifies the magnetic field on the scale of the fastest resolved eddies, which corresponds to our cell sizes. In a fluctuating small-scale dynamo, the scale of peak magnetic energy shifts to larger scales after the smaller scales have saturated.

\item The magnetic coherence length is further amplified by galactic winds and ram pressure stripping, which adiabatically expand the field and transport it into the ICM, as shown in Fig.~\ref{fig:a}.

\item By the time exponential amplification of the magnetic field develops in the ICM (at around $z\approx5.5$), the magnetic coherence length has already substantially grown to $\lambda_B\approx5~$kpc, as can be inferred from the power spectra in Fig.~\ref{fig:powerspec}, so that the dynamo in the ICM operates on these scales. The ICM particle mean free path at these high redshifts is smaller in comparison to nearby clusters due to the higher densities in the early Universe. The particle mean free path in the ICM, assuming it to be fully ionized, can be calculated as \citep{1965RvPP....1..205B}
\begin{align}
    \lambda_\mathrm{mfp} = \frac{3 \sqrt{2}}{4 \sqrt{\pi}} \frac{k_\mathrm{B}^2 T_\mathrm{i}^2}{\ln\Lambda_\mathrm{i} n_\mathrm{i} e^4} .
    \label{equ:mfp}
\end{align}
Here, $T_\mathrm{i}, n_\mathrm{i}, \ln\Lambda_\mathrm{i}$ are the ion temperature, number density, and Coulomb logarithm, respectively. The resulting particle mean free path in the ICM is shown in Fig.~\ref{fig:MFP} for $z<5.5$ (the blue curve shows the median and the 25$^\rmn{th}$ and 75$^\rmn{th}$ percentiles). Clearly, it is significantly smaller than the magnetic coherence length (shown in orange), which implies that the ICM dynamo remains in the fully collisional regime until $z = 0$. This justifies the use of the MHD approximation for modeling a dynamo in the ICM.
\end{enumerate}

This presents an alternative view to the plasma dynamo \citep{2018ApJ...863L..25S, 2019JPlPh..85a9014S,2020JPlPh..86e9003S,2022PNAS..11919831Z,2024ApJ...960...12Z}, where the magnetic seed field in the ICM is generated by plasma instabilities. In this scenario, large-scale turbulence in a weakly collisional plasma induces pressure anisotropies, triggering the Weibel instability, which creates magnetic seed fields. Kinetic instabilities, such as the firehose and mirror instabilities, transport these fields to larger scales. These instabilities also provide effective viscosity at Larmor radius scales, where the small-scale dynamo can then further amplify the magnetic field. Future research is needed to demonstrate a fully self-consistent computation of the plasma dynamo on astrophysically relevant scales. Additionally, further work is required to identify observational consequences that may differ from the proposed scenario.

\section{Mock observations}
\label{sec:6}

\begin{figure*}
  \includegraphics[width=\textwidth]{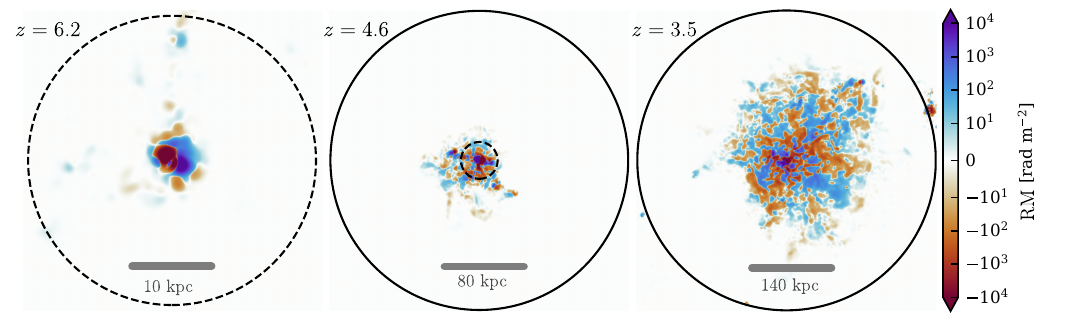}
  \caption{SKA mock Faraday RM observations of Halo3 for three different redshifts. The panels have sidelengths $40~\mathrm{kpc}$ (left) and $2 R_{200}(z)$ (middle and right), and a projection depth of $2R_{200}$. The virial radius $R_{200}(z)$ is indicated with the solid circle. The central part of radius $20~\mathrm{kpc}$, indicative of the BCG, is shown with the dashed circle. Observations of the magnetized (proto-)BCG are possible as early as $z=6.2$, where the cluster center shows high RM values, reflecting the interaction between AGN feedback and the magnetized ICM. From $z=4.6$ onwards, the entire cluster center and the magnetic field morphology can be studied with a dense enough grid of background and embedded polarized sources. At $z\leq3.5$, the magnetic field can be studied out to larger radii using Faraday rotation.}
  \label{fig:RM}
\end{figure*}

We simulate Faraday RMs at high redshifts. Our main finding is that magnetic fields are intrinsically linked to galaxy clusters as they form, originating in early galaxies and growing through subsequent mergers. Understanding these processes at high redshifts and in proto-BCGs is crucial and will be an active research topic of future observations.

Numerous radio observations of high redshift galaxies and protoclusters show that their magnetic field strength are comparable to that observed in nearby clusters \citep{2021NatAs...5..268D, 2022ApJ...937...45A, 2023A&A...676A..29C, 2024ApJ...961..120C, 2023A&A...675A..51D}. However, precise field strength estimates from radio brightness observations are challenging due to uncertainties associated with the equipartition assumption in galaxy clusters, i.e., whether the ratio of cosmic ray proton-to-magnetic energy density is indeed unity and whether the radio-emitting electrons constitute $1\%$ of the cosmic ray protons, as observed in the Milky Way \citep{2023A&ARv..31....4R}. Faraday rotation describes the effect of rotation of the plane of polarization in a magnetized plasma due to its birefringent property. In such a plasma, the rotation angle is given by $\phi=\phi_0 + \mathrm{RM}\, \lambda^2$, where $\phi$ and $\phi_0$ are the measured and intrinsic polarization angle, respectively, and $\lambda$ is the wavelength of radiation. The RM is given by \citep{2002ARA&A..40..319C} 
\begin{align}
    \mathrm{RM} = \frac{e^3}{2 \pi m_\mathrm{e}^2 c^4} \int_0^L \frac{n_\mathrm{e} \bm B \bcdot \mathrm{d} \bm l}{z+1}.
    \label{equ:RM}
\end{align}
Here, $e$ represents the electron charge, $m_\mathrm{e}$ the electron mass, $c$ the speed of light, $n_\mathrm{e}$ the number density of free electrons, $\bm B$ the magnetic field, and $L$ the physical distance from the radio source to the observer. These quantities are a direct output of our simulation, hence providing the possibility to create more accurate mock observations in comparison with e.g., radio maps that would require information about cosmic rays that are not simulated. Knowing the electron density along the line of sight (e.g., through thermal X-ray observations) enables us to infer the magnetic field strength, although this is however complicated by the non-monotonicity of the RM along the line of sight for a turbulent magnetic field and an inhomogeneous electron distribution \citep{2023arXiv230412350H}.

For our mock observations, we choose to mimic the Square Kilometre Array (SKA) due to its high angular resolution and sensitivity. SKA will provide an angular resolution of $\sim 0.002-0.04$'' at $0.5 - 10~\mathrm{GHz}$ \citep{Godfrey_2012}. We generate RM maps using a filter that smooths the image by performing a convolution with a Gaussian kernel with a standard deviation equivalent to a $0.04"$ resolution. We compute the angular size of the galaxy cluster using the angular distance between the cluster redshift and the observer at $z=0$ using the cosmology that we used to create the simulation introduced in Section~\ref{sec:pico}. For our adopted ISM model \citep{2003MNRAS.341.1253H}, we calculate the free electron number density following \cite{2018MNRAS.481.4410P}, where for star-forming regions, the fraction of the cold phase contributing to the density is subtracted. Additionally, we assume that the line-of-sight integral of RM is completely dominated by the virial region of the cluster. A contribution of |RM| of $10~\mathrm{rad~m}^{-2}$ corresponding to limiting values expected from the |RM| contributions of the Milky Way \citep{2018MNRAS.481.4410P} is not included (see also \citealt{2022ApJ...937...45A} who find a galactic foreground contribution of $13~\mathrm{rad~m}^{-2}$).

In Fig.~\ref{fig:RM}, we present the results of our RM mock observations for the SKA at three different redshifts $z=6.2, 4.6, 3.5$. The RM projections have sidelengths and projection depths of $2 R_{200}$. We indicate the virial radius (solid line) and the center as an inner circle with radius $20~\mathrm{kpc}$ (dashed line). A resolution of $0.04"$ enables the observation of the center of the simulated cluster, where the proto-BCG is active, at $z=6.2$. This results in strong signals of $\approx 10^4 ~ \mathrm{rad~m}^{-2}$ in the very center and in lower signals $\approx 10^2 ~ \mathrm{rad~m}^{-2}$ in a filamentary structure towards the top of the image. As redshift decreases, the signal strength grows while remaining nearly constant at the center. The increasing cluster size at lower redshifts allows for a more detailed study of the magnetic field's morphology and the identification of merging galaxies, such as those near the virial radius in the top right of the image. Faraday RM observations conducted by LOFAR of the BCG of the spiderweb galaxy cluster at $z=2.2$ demonstrate how the central AGN magnetizes the BCG's circumgalactic medium. Our mock observations agree with their actual observational RMs of $\sim 1500~\mathrm{rad~m}^{-2}$ in the source frame \citep{2022ApJ...937...45A}.

However, the effectiveness of such experiments is constrained by the density of background sources available for RM measurements. For SKA, we expect to reach $60~\mathrm{sources}/\mathrm{deg}^2$ \citep{2023MNRAS.526..836J}, which translates to $0.003$, $0.011$, $0.027$ sources per image for our $z=6.4$, $4.6$, $3.5$ mock observations respectively. Nevertheless, radio emission can also arise from the cluster-intrinsic sources in form of radio lobes from AGNs and radio relics/shocks \citep{2019SSRv..215...16V}. Radio shocks are thought to emerge from merger shocks that reaccelerate cooled cosmic ray electrons \citep{2013MNRAS.435.1061P,2024ApJ...961...15N}. Hence, instead of using a sparsely populated set of point sources to deduce the Faraday RM properties, we can use these diffuse and extended radio sources that continuously illuminate the central and peripheral cluster regions with polarized radio emission.

\section{Discussion and conclusions}
\label{sec:discussion}

\subsection{Comparison to observations and previous work}

Verifying these findings through observations is challenging. Magnetic field strength estimates from radio brightness rely on the uncertain equipartition assumption \citep{2023A&ARv..31....4R}, while estimates from Faraday RM data require precise electron density along the line of sight. Additionally, fluctuating magnetic field directions can cause cancellation effects. However, this issue can be addressed with high-quality RM data, which enables the derivation of the magnetic power spectrum using second-order Faraday RM statistics \citep{2005A&A...434...67V,2011A&A...529A..13K}. Despite the large uncertainties, radio observations have already provided some support for our findings. Clusters at approximately $z \approx 1$ show magnetic field strengths comparable to those in nearby clusters \citep{2021NatAs...5..268D, 2023A&A...675A..51D}. Additionally, observations of high-redshift protoclusters with redshifts in the range $4.25 > z > 2.2$ have revealed a magnetized ICM surrounding the BCG \citep{2022ApJ...937...45A, 2023A&A...676A..29C, 2024ApJ...961..120C}. Notably, Faraday RM measurements of a protocluster at $z=2.2$ yield RM values similar to those obtained from our mock observations of the radiative cluster at $z=3.5$ (cf.\ Fig.~\ref{fig:RM} and \citealt{2022ApJ...937...45A}). Moreover, we anticipate that the future capability of the SKA will enable RMs for massive proto-BCGs as early as $z=6.4$, thus providing an opportunity to test our predictions of early magnetic field growth.

However, the discrepancy between the magnetic Prandtl number in our simulations ($P_\mathrm{m} \approx 1$) and the real Prandtl numbers in the ICM ($P_\mathrm{m} \approx 10^{29}$, see e.g., \citealt{1993MNRAS.263...31T}; \citealt{2022hxga.book...56K}, Chapter 2.6.3) makes our simulation setup very different from reality. This discrepancy mainly influences the scale of magnetic field folding. Nevertheless, it has been shown that a $P_\mathrm{m} \approx 1$ dynamo behaves comparably to a large Prandtl number dynamo in growing the magnetic field, as the growth rate depends on the eddy turnover rates at the viscous scales (which is shown in e.g., \citealt{2004ApJ...612..276S}).

Our simulation methodology and findings are consistent with those of \cite{2015MNRAS.453.3999M}, who compared magnetic field evolution of a radiative simulation to a non-radiative simulation in a cosmological box. We expanded on their work with our analysis by employing a higher resolution, an updated galaxy formation model, and an extended analysis. Consistent with our findings, they reported faster growth and saturation in the radiative simulation, although they did not analyze the physical reason for it. Importantly, their amplification and saturation occurred at lower redshifts, indicating lower growth rates, which can be attributed to their lower resolution. Both our findings and those of \cite{2015MNRAS.453.3999M} underscore the significance of considering galaxy formation physics when investigating magnetic fields in galaxy clusters.

\cite{2022ApJ...933..131S} conducted a cosmological, non-radiative simulation of a cluster similar to our Halo3 but with half its mass. Despite using a resolution ten times better in mass than ours, they reported lower growth rates, with their magnetic field saturating at around $z \approx 2$. In contrast, \cite{2024ApJ...967..125S} simulated the same cluster at a resolution 20 times higher in mass than ours, finding that the magnetic field saturated at around $z = 4$, which is slightly earlier than our radiative simulation results. This suggests that including radiative physics can accelerate magnetic field growth, compensating for lower resolution to a certain extent, but not indefinitely.

In an ideal scenario of a small-scale dynamo, the saturation is dependent on the magnetic Prandtl number with $\mathrm{P_m} \gg1$ yielding higher ($43.8 - 1.3 \%$) saturation values of the magnetic-to-turbulent energy density and $\mathrm{P_m} \ll 1$ yielding lower ($2.43 - 0.135 \%$) saturation values \citep{2015PhRvE..92b3010S}. Our simulations, however, exhibit saturation levels of $\sim 10 \%$ of the total kinetic energy in the center of mass frame in the central region. These values align with observations \citep{2022SciA....8.7623B}, and they are also consistent with simulations of isotropic, homogeneous turbulence conducted by \cite{2012PhRvL.108c5002B}, where saturation values of approximately \mbox{4--5\%} were found. Similar agreement is observed in the simulations by \cite{2014MNRAS.445.3706V}, who explored the small-scale dynamo in galaxy clusters and reported an efficiency of kinetic turbulent energy conversion to magnetic energy of the order of $0.3$, still within the same order of magnitude as our results, albeit slightly higher. Also \cite{2024ApJ...967..125S} find saturation values of $0.1$ in their halo, which is similar in mass and size to our halo. In contrast, simulations of disk galaxies demonstrate much higher saturation values, approaching nearly $10$--$100\%$ \citep{2020MNRAS.498.3125P, 2021MNRAS.506..229W,2022MNRAS.515.4229P}. As \cite{2024MNRAS.528.2308P} showed, these high saturation values are linked to disk galaxies, whereas dwarf galaxies with more spherical shapes exhibit lower saturation levels. They attribute this difference to the presence of a mean-field rotational dynamo in disks, which is not present in objects without ordered rotation. 

Our results indicate that the BCG in the cluster center and merging galaxies play a pivotal role in enriching the ICM with magnetically pre-processed gas. However, the specific causes of gas loss from galaxies remain unclear. Our findings reveal that most galaxies entering the cluster arrive without self-bound gas. Moreover, we propose that AGN feedback may not significantly expel gas before $z=4$. Stellar feedback appears to assume a more prominent role at higher redshifts in the initial galaxies, but whether it generates sufficiently robust stellar winds extending beyond the galaxies' virial radii remains uncertain. Ram pressure stripping is likely to be a prevalent phenomenon in both the host cluster and pre-merging substructures. Our findings are consistent with those of \cite{2023arXiv231003083K} who explored the causes of galaxy quenching in the TNG simulation. Their results suggest that AGN feedback and ram pressure stripping contribute to galaxy quenching, with most galaxies in cluster vicinities being quenched at around $z \approx 3$. Due to various uncertainties in feedback mechanisms and high-redshift galaxy gas dynamics, which cannot be fully addressed within current galaxy formation models, we conclude that understanding the precise mechanisms responsible for galaxy quenching may not be necessary for concluding how the ICM is magnetized.

\subsection{Conclusions}

In this study, we analyzed two types of cosmological galaxy cluster simulations. The first, the `radiative' simulation, includes the galaxy formation model used in the IllustrisTNG simulations, while the `non-radiative' simulation focuses only on gravity and MHD. We examined the evolution of the magnetic field strength in four clusters, to account for sample variance and ensure robustness. In the remainder of our paper, we focused on the most massive cluster, Halo3, for detailed analysis, though the same qualitative results were found in the other clusters. Our analysis shows that the radiative simulation exhibits faster magnetic field amplification and saturation, as seen in Fig.~\ref{fig:B_time}. We also identified a connection between magnetic field evolution in the cluster and in galaxies. Our key findings are as follows:
\begin{itemize}

\item Cluster magnetic fields are intrinsically linked to galaxy clusters and their formation process. Initially, the magnetic field experiences amplification in the first galaxies that are formed in the cluster, which eventually merge to form the BCG. Subsequently, cluster mass and magnetic field grow with accretion of galaxies (Fig.~\ref{fig:timeseries}).

\item Magnetic fields grow fast within the forming galaxies at early times ($z\approx 8$) due to a fluctuating small-scale dynamo that is excited by super-sonic, compressible (Burgers) turbulence induced by gravitational infall of cold gas (Fig.~\ref{fig:feedback_projection}). We have the following evidence supporting the existence of such a dynamo: 1) the magnetic field of the ISM grows exponentially in time (Fig.~\ref{fig:B_time}), 2) a power spectrum analysis reveals a Kazantsev scaling in kinematic phase of the dynamo on large scales, and after saturation of the small-scale field, we observe the growth of the magnetic coherence length (Fig.~\ref{fig:powerspec}), and 3) the fitted exponential growth rates agree with the theoretically expected rates of Burgers turbulence when we adopt our numerical Reynolds number (Fig.~\ref{fig:b}). This growth is already sufficient to match observational inferences and could be even faster for realistic Reynolds numbers in the ISM. Eventually, this leads to a saturated magnetic field in the ISM at around $z \approx 4$. 

\item The magnetized gas in the BCG is transported to the surrounding ICM via turbulence that is injected by stellar feedback and the interaction with accreted gas, and (to a lesser extend) AGN feedback (Figs.~\ref{fig:feedback} and \ref{fig:feedback_projection}).

\item Later on merging galaxies are responsible for enriching the ICM with the galactic magnetic field by means of galactic outflows driven by SN and AGN feedback as well as ram pressure stripping. Pure adiabatic expansion of the magnetic field in star formation regions to ICM densities could explain a large fraction of the ICM magnetic field strength (see Fig.~\ref{fig:a}). However, this overestimates the measured field strengths, as this simplified procedure neglects mixing losses and magnetic fields trapped in stars or retained in the ISM. This demonstrates that ``pre-processing'' \citep{2004PASJ...56...29F}, commonly associated with metal enrichment, likely also plays a significant role in the magnetic enrichment of the ICM in merging structures, which also manifests as a metallicity-magnetic field strength correlation in the ICM (Fig.~\ref{fig:Phase}).

\item Cluster mergers inject incompressible (Kolmogorov) turbulence (Fig.~\ref{fig:powerspec}) that triggers a fluctuating small-scale dynamo in the entire ICM at lower redshifts ($z\lesssim 5.5$). This dynamo utilizes the pre-magnetized plasma expelled from galaxies as seed fields for further amplification as we demonstrate by the following points: 1) a power spectrum analysis of the kinetic and magnetic energy density in the ICM reveals a Kazantsev scaling on large scales and shows an increasing magnetic coherence length after saturation on small scales (Fig.~\ref{fig:powerspec2}), 2) this dynamo is able to maintain an approximately constant volume-averaged field strength until $z=0$ (Fig.~\ref{fig:B_time}) and counteracts the expected decrease of magnetic field strength over time as a result of mixing of accreted gas with smaller magnetization, giving rise to a constant magnetic-to-kinetic energy ratio after $z \approx 3.5$ (Fig.~\ref{fig:Energies}), and 3) the correlation of $B$ and gas density follows the relation of self-gravitating clouds and cluster observations (Fig.~\ref{fig:Phase}).
    
\item Our results suggest that magnetic field strengths in clusters at redshift $z=3.5$ should be comparable to those in nearby clusters (Fig.~\ref{fig:B_time}). Moreover, in the central galaxy, magnetic field strengths should be larger than 10~$\upmu$G as early as $z=5.5$ (Figures~\ref{fig:B_time} and \ref{fig:timeseries}). 

\item We demonstrated that the magnetic field observed today in the \emph{weakly collisional} ICM is always amplified on \emph{collisional} scales: first, in the collisional ISM during the epoch of BCG assembly at $z\approx8$ and later on, in the ICM on the magnetic coherence scale, which is always larger than the particle mean free path (Fig.~\ref{fig:MFP}). Hence, this justifies the use of MHD for studying the cluster dynamo and facilitates and by and large bypasses the extraordinary challenge to explain magnetic field growth in a weakly collisional plasma from the tiny seed values \citep{2020JPlPh..86e9003S,2024ApJ...960...12Z}.
\end{itemize}

A major outcome of our study is the realization that the magnetic field within the cluster experiences its initial amplification at high redshifts within the protocluster BCG. This amplification is attributed to compressive turbulence generated by gravitational infall of cold gas, which excites a small-scale dynamo. While previous research has theoretically explored how compressive turbulence can enhance magnetic fields in galaxies, this phenomenon has not yet been seen in cosmological simulations of galaxy clusters. Therefore, it is essential to rigorously test this outcome and determine its robustness for different galaxy formation models in the future.

Theoretical studies of the magnetic field evolution in galaxy clusters align with the forthcoming or ongoing telescope surveys. The James Webb Space Telescope (JWST) can illuminate the formation and morphology of high-redshift protoclusters \citep{2023MNRAS.526L..56F, 2023MNRAS.526L..47M}. The recently launched X-Ray Imaging and Spectroscopy Mission (XRISM) is primed to study ICM turbulence in nearby clusters with exquisite detail \citep{2023MNRAS.518.2954D, 2023HEAD...2010129M}. Radio surveys like the LOFAR Two-metre Sky Survey (LoTSS) \citep{2017A&A...598A.104S, 2021NatAs...5..268D, 2023A&A...675A..51D} and the upcoming Square Kilometer Array (SKA) \citep{2018A&A...617A..11G} promise to explore the interaction between cosmic rays and magnetic fields within the ICM across a large range of redshifts. Hence, the motivation for theoretical studies employing simulations remains compelling.

\begin{acknowledgements}
      We acknowledge support by the European Research Council under ERC-AdG grant PICOGAL-101019746 and by the German Science Foundation (DFG) through the Research Unit FOR-5195. TB gratefully acknowledges funding from the European Union’s Horizon Europe research and innovation programme under the Marie Skłodowska-Curie grant agreement No 101106080. JW acknowledges support by the German Science Foundation (DFG) under grant 444932369.       RW acknowledges funding of a Leibniz Junior Research Group (project number J131/2022).
\end{acknowledgements}

\noindent \textbf{Software} \\

\noindent    We use Paicos \citep{Berlok2024} to load and process the data. Paicos is a parallelized object-oriented Python package for analysis of simulations performed with AREPO \href{https://github.com/tberlok/paicos}{https://github.com/tberlok/paicos}. 

%
%
\bibliographystyle{aa}
\bibliography{bib}

\begin{appendix}

\section{Spherical profiles and Fourier transforms}
\label{sec:density}

\subsection{Spherical profiles}
\label{sec:profiles}

\begin{figure*}[h!]
  \includegraphics[width=\textwidth]{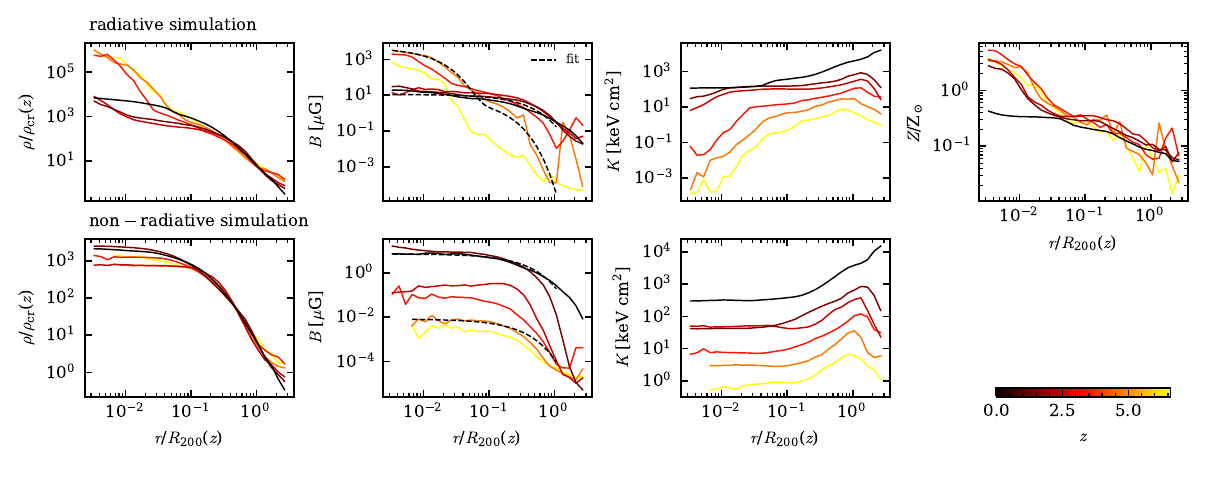}
  \caption{We present spherically averaged radial profiles of density, magnetic field, entropy and metallicity for the radiative (upper panel) and the non-radiative simulation (lower panel). We show 6 different redshifts spanning a range $z=0-6.5$ and a radius $10^{-3} R_{200}<r<R_{200}$. We fit an exponential to the magnetic field profiles at $z=4.5$ and $z=0$ in the range $0.003 R_{200}<r<R_{200}$. The radiative simulation shows a similar evolution in all profiles: at high redshifts, core values are very high. With decreasing redshift, mergers and feedback processes distribute the enhanced core values over larger radii. Both, metallicity and magnetic fields are results of pre-processing from galaxies. The profiles of the magnetic field strength experience a stronger increase at the outer radii in comparison to the metallicity profiles, which shows that a dynamo is operating on top of the pre-processed gas.}
  \label{fig:Profiles}
\end{figure*}

We examine the co-evolution of magnetic field, density and metallicity. Figure~\ref{fig:Profiles} shows spherically averaged radial profiles of Halo3 for density, magnetic field strength, entropy, and metallicity in radiative (upper panel) and non-radiative (lower panel) simulations for 6 different redshifts.
In the second panel column of Fig.~\ref{fig:Profiles}, we present spherically averaged, volume-weighted root mean square magnetic field strength profiles. We fit single- or double-exponential profiles to our simulated magnetic profiles, depending on the requirement at a specific redshift: 
\begin{align}
    B (r) &= B_{0} \mathrm{e}^{ - k_0 r}, \label{eq:single_exp} \\
    B (r) &= B_{1} \mathrm{e}^{ - k_1 r} + B_{2} \mathrm{e}^{ -k_2 r},
    \label{eq:double_exp}
\end{align}
where $B_i$ ($i\in\{0,1,2\}$) are the magnetic field strengths at $r=0$ and $k_i^{-1}$ are the characteristic scales of the corresponding exponential profiles and have units of $\mathrm{kpc}$. We fit both of our simulations at $z=0$ and $z=4.5$ in the range $0.003 R_{200}<r<R_{200}$ and report the fit values in Table~\ref{table:2}. Note that at $z=0$, in the saturated dynamo regime, the magnetic field profile scales with the square root of the density profile (see Sect.~\ref{ssec:PSDig}), which could explain the slight deviations between the fitted and measured curves. In the non-radiative run, the growth is smooth, while in the radiative simulation, magnetic fields increase exponentially, with exceptionally high strengths at the center at $z=3.5$ (these were not seen previously when averaging over a larger volume).

The right panel of Fig.~\ref{fig:Profiles} displays metallicity profiles exhibiting a similar trend: elevated values in the center at high redshift and a smoother distribution at lower redshifts. The transformation from cuspy to smooth occurs around $z=3.5$. At $z=0$, the distribution becomes even flatter, in comparison to the magnetic field profile. 

Gas transport mechanisms are still a subject of investigation. In our model, stellar and AGN feedback contribute (especially in destroying the cool core), but mergers appear to be the primary drivers, injecting substantial energy that stirs the gas. Mergers introduce pre-enriched gas to the outskirts, flattening the magnetic and metallicity profiles (one merger can be seen as an extended bump in the magnetic field and metallicity profiles at $z=3.5$ after which the profiles are flattened). The magnetic profiles experience a steeper increase in the outskirts in comparison to the metallicity profiles. This shows how the dynamo acts on top of the pre-enriched gas. The similarity between the density and magnetic profiles, i.e., the enhanced cores at high redshifts and the subsequent flattening, shows the contributions of adiabatic compression to the magnetic evolution. Also here, the magnetic field experiences a stronger increase in the outskirts, which is because of the small-scale dynamo that additionally amplifies the field.

Our study of pre-enrichment aligns with previous research, emphasizing the significant role of mergers in shaping these profiles. Notably, \cite{2018MNRAS.474.2073V}, employing the same TNG galaxy formation model, report analogous findings. They find that by $z=0$, $80 \%$ of all metal in the ICM are accreted. 

\begin{table*}
\caption{Redshift, simulation and fit parameters for the magnetic exponential radial profiles in Eqs.~\eqref{eq:single_exp} and \eqref{eq:double_exp}.}
\label{table:2}
\centering
\begin{tabular}{c c c c c c c c}
\hline\hline
$z$ & Simulation & $B_0$ & $B_1$ & $B_2$ & $k_0$  & $k_1$ & $k_2$\\
   & & $[\upmu\mathrm{G}]$ & $[\upmu\mathrm{G}]$ & $[\upmu\mathrm{G}]$ & $[\mathrm{kpc^{-1}}]$ & $[\mathrm{kpc^{-1}}]$ & $[\mathrm{kpc^{-1}}]$ \\
\hline
   4.5 & radiative  & - & 4470.571 &  6.122 & - & 0.685 & 0.075 \\
   4.5 & non-radiative & 0.008 & - & - & 0.029   & - &  - \\
   0 & radiative  & 10.997 & -     & -  & 0.001 & - & -  \\
   0 & non-radiative & 7.199 & -    & -  & 0.001 & - & - \\
\hline
\end{tabular}
\end{table*}

\subsection{Fourier transform and power spectra}
\label{sec:Fourier_transform}

As we have seen, the magnetic profiles in the radiative simulation at $z=0$ ($z=4.5$) follow a (double) exponential profile. In order to compute the power spectrum, we need to first take the Fourier transform of a single exponential profile:
\begin{align}
\mathcal{F}\left(\mathrm{e}^{-k_0 r} \right)(k) 
&= \int_{-\infty}^{\infty} \mathrm{d}^3 x~  \mathrm{e}^{-k_0 \sqrt{x^2 + y^2 + z^2}} \mathrm{e}^{- i \bm k \bm x} \nonumber\\ 
&= 4 \pi \int ^{\infty} _0 \mathrm{d}r r^2 \mathrm{e}^{-k_0 r} \frac{\mathrm{sin} \left(kr \right)}{kr}  
= \frac{8\pi k_0 }{\left[ k^2 + k_0^2 \right]^2},
\label{equ:lorentz}
\end{align}
where $r = \sqrt{\bm x^2}$ and $k = \left| \bm k \right|$. The 1D power spectrum $P(k)$ is then obtained via:
\begin{align}
P(k) &= 4 \pi  k^2 P_{3\mathrm{D}}(k)\propto k^2 \left| \mathcal{F}\left(B_0 \mathrm{e}^{-k_0 r} \right)(k) ) \right|^2 = \frac{\left( B_0 8\pi k_0 k \right)^2}{\left[ k^2 + k_0^2 \right]^4}
\label{equ:lorentz2}
\end{align}
In the case of a double-exponential profile, the power spectrum is given by the absolute square of the Fourier transform of Eq.~\eqref{eq:double_exp}.

\end{appendix}

\end{document}